\documentclass[aps,prd,twocolumn,nofootinbib,
superscriptaddress,floatfix,preprintnumbers]{revtex4-2}

\usepackage{amsmath,amssymb,amsfonts,bm}
\usepackage{graphicx}
\usepackage{dcolumn}
\usepackage[dvipsnames]{xcolor}
\usepackage{slashed}
\usepackage[normalem]{ulem}
\usepackage{tikz}
\usepackage[colorlinks=true, linkcolor=blue, urlcolor=Blue, citecolor=Mulberry]{hyperref}
\usepackage{booktabs}
\usepackage{orcidlink}
\allowdisplaybreaks
\newcommand{\beq}{\begin{equation}\begin{aligned}}
\newcommand{\eeq}{\end{aligned}\end{equation}}

\newcommand{\mchi}{m_\chi}

\begin{document}
%%%%%%%%%%%%%%%%%%%%%%%%%%%%%%%%%%%%%%%%%%%%%%%%%%%%%%%%%%%%
% Title
%%%%%%%%%%%%%%%%%%%%%%%%%%%%%%%%%%%%%%%%%%%%%%%%%%%%%%%%%%%%
\title{Multipolar Dark Matter Freeze-out in an Early Matter-Dominated Universe}

\author{Debajit Bose\,\orcidlink{0000-0001-8594-8885}}
\email{debajitbose550@gmail.com}
%\email{debajitbose550@gmail.com}
\affiliation{Centre for High Energy Physics, Indian Institute of Science, C. V. Raman Avenue, Bengaluru 560012, India}

\author{Prolay Chanda\,\orcidlink{0000-0002-3940-6062}}
\email{prolay.chanda@tifr.res.in}
\thanks{Corresponding author.} 
\affiliation{Department of Theoretical Physics, Tata Institute of Fundamental Research, Mumbai 400005, India}

\author{Suvam Maharana\,\orcidlink{0000-0001-9313-4064}}
\email{suvam.maharana\_528@tifr.res.in}
\affiliation{Department of Theoretical Physics, Tata Institute of Fundamental Research, Mumbai 400005, India}

\author{Poulami Mondal\,\orcidlink{0000-0002-4164-6057}}
\email{poulami.mondal@tifr.res.in}
\affiliation{Department of Theoretical Physics, Tata Institute of Fundamental Research, Mumbai 400005, India}

%%%%%%%%%%%%%%%%%%%%%%%%%%%%%%%%%%%%%%%%%%%%%%%%%%%%%%%%%%%%
% Abstract
%%%%%%%%%%%%%%%%%%%%%%%%%%%%%%%%%%%%%%%%%%%%%%%%%%%%%%%%%%%%
\begin{abstract}
The relic abundance of thermal dark matter depends not only on its particle interactions but also on the expansion history of the early Universe. We study the freeze-out of fermionic dark matter interacting with the Standard Model through higher-dimensional electromagnetic operators in an early matter-dominated cosmology. In particular, we consider magnetic-dipole, electric-dipole, anapole, and charge-radius interactions and compute the interaction strengths required to reproduce the observed dark matter relic abundance, accounting for the subsequent entropy injection from the decay of a long-lived heavy field. We compare the resulting parameter space with that obtained in the standard radiation-dominated freeze-out scenario and confront it with current constraints from direct-detection experiments and solar-neutrino searches. We find that the entropy dilution associated with an early matter-dominated epoch significantly reduces the interaction strength required to obtain the observed relic abundance, thereby opening regions of parameter space that are excluded within the conventional cosmological history.  
\end{abstract}

\preprint{TIFR/TH/26-20}

\maketitle

\section{Introduction}

The nature of particle dark matter (DM) remains one of the most intriguing puzzles in modern physics. In the absence of a clear detection  thus far, the characterization of possible DM candidates is essentially governed by two fundamental elements, namely the precise nature of the early Universe cosmology and the microscopic dynamics dictating the interactions of the DM candidates among themselves  as well as with the Standard Model (SM) particles.

On the cosmological side, the Universe is generally understood to have been radiation-dominated during Big Bang Nucleosynthesis (BBN)~\cite{Allahverdi:2020bys}. Owing to our limited knowledge of the pre-BBN era, a common assumption is, therefore, that radiation domination persisted from post-inflationary reheating until BBN. However, alternative thermal histories of the early Universe remain viable, since there is no direct evidence requiring the Universe to have been radiation-dominated throughout the entire period between inflationary reheating and BBN. The possibility of such \emph{non-standard} cosmologies has elicited extensive studies of their attendant consequences, especially in the context of DM physics~\cite{Kamionkowski:1990ni,Giudice:2000ex,Moroi:1999zb,Lyth:1993zw,Gelmini:2006pq,Gelmini:2006pw,Acharya:2009zt,Kane:2015qea,Davoudiasl:2015vba,Randall:2015xza,Co:2015pka,Dror:2016rxc,Banerjee:2022fiw,Khlopov:1980mg,Datta:2025vyu,Mansour:2026sdx}.

The weakly interacting massive particle (WIMP) paradigm has historically provided a compelling DM candidate, since stable particles with weak-scale masses and weak-strength interactions naturally produce the observed relic abundance via thermal freeze-out~\cite{Kolb:1990vq}.While DM freeze-out calculations are typically performed under the assumption of radiation domination, freeze-out may instead have taken place during an era in which the dominant energy density redshifted differently from radiation, thereby affecting the DM freeze-out dynamics.  DM freeze-out during an early matter-dominated era is one of such scenarios that has been studied in ref.~\cite{Hamdan:2017psw}, following the original motivation of ref.~\cite{Kamionkowski:1990ni}. It assumes the presence of a heavy field that comes to dominate the energy density of the Universe, leads to an early period of matter domination, and subsequently decays into radiation prior to BBN, thereby injecting entropy into the thermal bath. An important consequence of an early matter-dominated era is that it allows for overproduction of DM at freeze-out, which gets diluted later due to the entropy injection. Equivalently, this allows for weaker annihilation cross-sections, which have been shown to reopen viable parameter space for otherwise strongly constrained DM freeze-out models~\cite{Hamdan:2017psw,Chanda:2019xyl,Chanda:2021tzi,Asadi:2021bxp}.

 As for the particle identity of DM and its interactions with SM particles, there exist a plethora of possibilities, most notably within the WIMP paradigm~\cite{Steigman:1984ac,Jungman:1995df,Bertone:2004pz,Steigman:2012nb,Arcadi:2017kky,Kundu:2019nqo}. What then serves to reduce the number of possibilities is the assumption of minimality. One such minimal, albeit interesting, possibility is that DM interacts only electromagnetically with the SM sector. The scenario in which DM carries an electric charge is strongly disfavored by cosmological observations~\cite{Xu:2018efh,Dvorkin:2019zdi}, astrophysical probes~\cite{chang:2018rso,Caputo:2019tms,Fiorillo:2024upk}, and direct detection experiments~\cite{Iles:2024zka}. Therefore, a well-motivated alternative is that DM is electrically neutral and yet interacts  electromagnetically  through higher-dimensional operators --- the leading ones being characterized by electromagnetic multipole moments~\cite{Zeldovich:1958, Pospelov:2000bq, Sigurdson:2004zp, Masso:2009mu, Barger:2010gv, Ho:2012bg, Kavanagh:2018xeh, Hambye:2021xvd, Ibarra:2022nzm}. For a Dirac fermion DM, these comprise (up to dimension six) the magnetic dipole moment, electric dipole moment, charge radius and anapole operators. In the case of a Majorana fermion DM, the only possible electromagnetic couplings are furnished by the anapole operator and its higher moments~\cite{Radescu:1985, Ho:2012bg}. 

However, previous studies of multipolar DM candidates have derived strong limits from direct detection~\cite{Masso:2009mu, Banks:2010eh, DelNobile:2014eta, Chatterjee:2022gbo, Ibarra:2022nzm, Bose:2023yll, Ibarra:2024mpq, Kumar:2025wsp}, astrophysical and cosmological observations~\cite{Geytenbeek:2016nfg, Lambiase:2021xcj, Bose:2023yll, Das:2025bnr}, as well as collider experiments~\cite{Fortin:2011hv, Chu:2018qrm, Biswas:2026ahs}.~When combined with the DM relic density requirement under the assumption of a radiation-dominated freeze-out scenario, these bounds leave the parameter space strongly constrained.
In this work, our key goal is to consider the case of matter-dominated freeze-out as discussed in the previous paragraphs and explore whether an alternative early cosmology can help evade the existing experimental bounds on the parameter space. 

This paper is structured as follows. In Sec.~\ref{sec:MDFO}, we discuss the matter-dominated freeze-out scenario and calculate the DM relic abundance after freeze-out and the subsequent entropy injection. In Sec.~\ref{sec:multipole_theory}, we mention several ways a neutral DM candidate can interact with the SM electromagnetically and derive the DM annihilation cross-section to the SM for such models. In Sec.~\ref{pheno}, we present the allowed parameter space, and discuss how an early matter-dominated era relaxes the experimental constraints on the multipolar DM models considered in this work. We finally conclude in Sec.~\ref{sec:conclusions}.

\section{Matter Dominated Dark Matter Freeze-Out}
\label{sec:MDFO}

In this section, we discuss the DM freeze-out mechanism in an early matter-dominated cosmology, originally proposed in ref.~\cite{Hamdan:2017psw}.~We elaborate on the differences between the matter-dominated freeze-out from the conventional radiation-dominated freeze-out mechanism; for example, the change in the expansion history and the presence of a heavy field that results in early matter-dominated era.  We present an analytic prescription for the Boltzmann analysis for calculating the freeze-out temperature and DM relic density. 

\subsection{General Setup and Dark Matter relic abundance}

Following the discussion of~\cite{Hamdan:2017psw}, we introduce a sufficiently long-lived heavy scalar $\phi$ that starts evolving as matter at a critical temperature $T_\star$ such that its energy density $\rho_\phi$ scales as $a^{-3}$, where `$a$' is the cosmological scale factor. Therefore, given $\phi$ is long-lived enough, there will be a matter dominated period in the early Universe. In the simplest scenario, $\phi$ decouples from the thermal bath so that $\phi$ starts behaving as matter at $T_\star\sim m_\phi$, where, $m_\phi$ is the mass of the heavy scalar \footnote{In general, if $\phi$ is decoupled from the thermal bath, $T_\star$ represents the Standard Model temperature at which $\phi$ becomes matter-like. Also, $\phi$ could be a fermion or a vector boson instead of a scalar. Importantly, $\phi$ is assumed to have no coupling with the DM at the tree-level.}. This early matter domination ends following the decay of the $\phi$ field and transitions to a radiation-dominated era. From here on, every quantity with subscript `$\star$' is evaluated at $T=T_\star$.

The fraction of energy density contained in radiation at $T_\star$ can be defined as follows
 \beq
r \equiv \frac{\rho_R + \rho_\chi}{\rho_R + \rho_\chi + \rho_\phi}\Bigr|_{T=T_\ast},
\eeq
where, $\rho_\chi, \rho_R$ is the energy density of the DM, $\chi$, and the SM thermal bath, respectively.

When $T>T_\star$, both the DM and $\phi$ behave as radiation along with the SM thermal bath. The energy density of the radiation-like component is determined as
 $(\pi^2/30)g_R T^4$, where, $g_R$ is the internal degrees of freedom in radiation.

The Hubble parameter is determined by the total energy budget of the Universe
\begin{equation}
H^2 = \frac{8\pi}{3 M_{\rm Pl}^2}\left(\rho_R + \rho_\phi + \rho_\chi\right),
\end{equation}
which, for practical purposes, we rewrite in the following form 
\beq
\frac{H^2}{H_\star^2}=\!\left[
\frac{g_\ast r}{g_\ast+g_\chi}\!\left(\frac{a_\star}{a}\right)^4
+(1-r)\!\left(\frac{a_\star}{a}\right)^3
+\frac{g_\chi r}{g_\ast+g_\chi}\!\left(\frac{a_\star}{a}\right)^4
\right].
\eeq

Here, the normalization factor is the Hubble parameter, $H_\star\, \equiv\, H(T_\star)\,\approx\, \sqrt{4\pi^3/45}g_{\ast}^{1/2}(T_\star)T_\star^2/M_{\rm Pl}$, because $g_\ast\gg g_\chi, g_\phi$, where, $g_\ast,~g_\chi,~g_\phi$ are the internal degrees of freedom of the SM, DM and $\phi$.

Assuming entropy conservation in the radiation bath prior to significant $\phi$ decay,
the scale factor $a$ and temperature $T$ are related as
\begin{equation}\label{eq:a-to-T}
\left(\frac{a_\star}{a}\right)
\simeq \left[\frac{g_\ast(T)}{g_\ast(T_\star)}\right]^{1/3} \frac{T}{T_\star},
\end{equation}
allowing the Friedmann equation to be expressed explicitly in terms of $T$.

\subsection{Modified Hubble Expansion and Freeze-Out Condition}

The freeze-out condition is defined by equating the DM annihilation rate per DM particle $\Gamma_{\rm ann,\chi}$, with the Hubble parameter
\begin{equation}
\Gamma_{\rm ann,\chi}  = H(T_{\rm f}),
\end{equation}
such that at a thermal bath temperature $T_{\rm f}$, the DM particles cease to annihilate and their abundance freezes out.

Thus, the freeze-out conditions depend strongly on the  expansion history of the Universe -- namely, the standard radiation-dominated universe ($H\propto T^2$), early matter-dominated universe ($H\propto T^{3/2}$)~\cite{Kamionkowski:1990ni,Hamdan:2017psw}, particle decay/ heating ($H\propto T^4$)~\cite{Scherrer:1985, Giudice:2000ex}, kination ($H\propto T^3$)~\cite{Spokoiny:1993kt}, etc.

We consider the simplest scenario of a purely matter-dominated Universe such that the decay of $\phi$ can be neglected when the DM freeze-out occurs. Apart from the different expansion rate in an early matter-dominated Universe, the entropy produced by the decay of $\phi$ dilutes the DM abundance produced at freeze-out, reducing the relic abundance by a factor $\zeta$,
\begin{equation}
\Omega_{\rm relic,\chi }= \zeta\,\Omega_{\rm f,\chi},
\end{equation}
which we call as the entropy dilution factor.

We define the decay temperature \(T_\Gamma\) as the point when the \(\phi\) decay width $\Gamma_\phi$ becomes comparable to the Hubble expansion rate $H(T_\Gamma)$. In the sudden decay approximation, the reheating temperature, $T_{\rm RH}$,\footnote{ Not to be confused with the inflationary reheating temperature.} corresponding to $\phi$ is given as $T_{\rm RH}\simeq\sqrt{\Gamma_\phi M_{\rm Pl}}$, and, the entropy dilution factor is given as~\cite{Randall:2015xza}
\begin{equation}\label{eq:zeta-TRH}
\zeta\equiv \frac{s_{\rm before}}{s_{\rm after}} \simeq \left(\frac{T_\Gamma}{T_{\rm RH}}\right)^3,
\end{equation}
where, $s_{\rm before}(s_{\rm after})$ is the entropy density before (after) $\phi$ decay. Using Eqs.~\eqref{eq:a-to-T}~and~\eqref{eq:zeta-TRH}, we can derive the following form  for $T_{\Gamma}$, relating $T_{\rm RH}$ and the dilution factor~\cite{Chanda:2019xyl}
\beq\label{eq:Tgamma-TRH}
T_{\Gamma} \simeq \left(\frac{45}{4\pi^3 g_\star (1-r)}\frac{T_{\rm RH}^4}{T_{\star}}\right)^{1/3}.
\eeq

\subsection{Dark Matter Yield and Relic Density}\label{sec:DM-relic density}

In the initially radiation dominated Universe, at a temperature $T_\star$ the $\phi$ field becomes non-relativistic, its energy density grows relative 
to radiation and eventually drives the Universe into matter domination. 
We denote by $T_{\rm MD}$ the temperature at which this transition occurs, 
defined as the point where the $\phi$ field contributes half of the total 
energy density. This temperature is related to $T_\star$ via~\cite{Chanda:2019xyl,Chanda:2021tzi}
\beq
T_{\rm MD}=\frac{1-r}{r}\left( \frac{g_\ast(T_\star)}{g_\ast(T_{\rm MD})} \right)^{1/3}T_\star.
\eeq

In both matter- and radiation-dominated cosmologies, the Hubble parameter 
scales inversely with time $t$. It is therefore convenient to adopt the 
parametrization
\beq
H = \frac{\nu}{t},
\eeq
with $\nu = 2/3$ and $\nu = 1/2$ for matter and radiation domination, 
respectively.

The Boltzmann equation for the DM relic abundance $Y_\chi$ is 
\beq
\frac{dY_\chi}{dt} = -\langle\sigma v\rangle s(T)(Y_{\chi}^2 - Y^2_{\rm \chi, eq}),
\eeq
where, $x=m_\chi /T$, $\langle\sigma v\rangle$ is thermally-averaged annihilation cross-section, $s(T)$ is the entropy density,  and the equilibrium abundance for DM is~\cite{Graesser:2011wi}
\beq\label{eq:Yeq}
Y_{\chi,\rm eq}&=bx^{3/2}e^{-x},
\eeq
where $ b = (45g_\chi)/(4\sqrt{2}\pi^{7/2}g_{\ast S})$ and $g_{\chi}$ is the DM internal degrees of freedom (2 for complex scalar, 4 for Dirac fermion). The DM annihilation cross-section can be expanded in a power series around
$x = 0$, 
\beq
\langle \sigma v \rangle = \sum_{n=0}^{\infty} \sigma_n x^{-n}.
\eeq
The DM abundance equation can, then, be reorganized as
\beq\label{eq:BE-eqn}
\frac{dY_{\chi}}{dx}= - \sum_n \lambda_n \, x^{-n-2} \, g_{\rm eff}^{1/2}
\left( Y_\chi^2 - Y^2_{\chi,\rm eq}\right),
\eeq
where the quantity $\lambda_n$ contains the cross-section and is defined as
\beq
\lambda_n=\nu \, \sqrt{\frac{4\pi}{45}}\, m_\chi M_{\rm Pl} \sigma_n,
\eeq
with $g_{\rm eff}$ being the effective internal degrees of freedom given as
\beq\label{eq:geff}
g_{\rm eff}^{1/2}\simeq g_{\star}^{1/2}(T_\star)\left(\frac{3\,T_\star (1-r)\, x}{4\, m_\chi}+ r\right)\left(\frac{T_\star (1-r)\, x}{m_\chi}+ r
\right)^{-3/2}.
\eeq
As we show in the App.~\ref{sec:derivation-FO}, an approximate analytic derivation of this differential equation provides the following form for the freeze-out abundance 
\beq\label{eq:MDFO-density}
\Omega_{\rm relic}^{\rm MD}
&=\zeta\,\Omega_{\rm f}^{\rm MD},\\
\Omega_{\rm f}^{\rm MD}h^2&=\frac{10^9}{M_{\rm Pl}}\left(\sum_n\frac{
\sqrt{g_*}\,(1-r)^{-1/2}\,x_\star^{1/2}\,\sigma_n\,{\rm GeV}}{(n+3/2)\,x_{\rm f}^{\,n+3/2}}\right)^{-1}.
\eeq
For reference, in the case of radiation-dominated freeze-out~\cite{Scherrer:1985zt} the relic density is given as
\beq\label{eq:RDFO-density}
\Omega_{\rm relic}^{\rm RD}h^2 = \Omega_{\rm f}^{\rm RD}h^2&\simeq\frac{10^9}{M_{\rm Pl}}\left(\sum_n\frac{\sqrt{g_*}\,\sigma_n\,{\rm GeV}}{(n+1)\,x_{\rm f}^{\,n+1}}\right)^{-1}.
\eeq
The freeze-out temperature for matter domination and radiation domination is given as follows
\beq\label{eq:xf-MD-RD}
x_{\rm f}^{\rm MD}& = \ln\left[\tilde{A}m_{\chi}^{3/2}T_{\star}^{-1/2}(1-r)^{-1/2}\sum_n(n+1)\sigma_{n}x_{\rm f}^{-n-1}\right],\\
x_{\rm f}^{\rm RD}& = \ln\left[\tilde{A}m_{\chi}\sum_n(n+1)\sigma_{n}x_{\rm f}^{-n-1/2}\right],\\
\eeq
where, the prefactor is defined as $$\tilde{A}= \frac{\sqrt{45}g_\chi}{\sqrt{32}\pi^3}g_{\star}^{-1/2}M_{\rm Pl}.$$ A detailed derivation of the above equations is given in App.~\ref{sec:derivation-FO}. We also discuss in Sec.~\ref{sec:results} the validity of sudden decay of $\phi$, an assumption that we made while deriving the DM relic density during matter domination.

\subsection{Consistency conditions}\label{sec:consistency1}

The freeze-out analysis presented before relies on several assumptions, particularly regarding the hierarchy of cosmological energy scales and the ordering of the relevant processes in the early Universe. In this section, we outline the cosmological consistency requirements necessary for the validity of our treatment. 

With the characteristic scales defined, the parameter space relevant for matter-dominated freeze-out is restricted by the following consistency conditions:
\begin{itemize}
\item DM freeze-out happens during the matter-dominated epoch, $
T_{\rm f} < T_{\rm MD} \lesssim T_\star$.
\item The reheating resulting from $\phi$ decays must occur before BBN, such that
$T_{\rm RH} > T_{\rm BBN} \sim 10\,{\rm MeV}.$, where, $T_{\rm BBN}$ is the characteristic BBN temperature~\cite{Hasegawa:2019jsa}.
% \item Freeze-out should occur before a major fraction of $\phi$ decays,
% $T_{\rm f} \gg T_\Gamma ,$
% ensuring that the freeze-out process happens during matter-domination.
% \item The decays of $\phi$ must also be negligible during freeze-out  so that the entropy of the radiation bath is approximately conserved,
% $T_{\rm f} \gg \, T_{\rm RH}.$
\item Freeze-out must occur well before the decay of $\phi$ becomes important. We therefore require $T_{\rm f} \gg T_\Gamma,T_{\rm RH}$
so that the matter-dominated freeze-out treatment remains applicable. For a continuous decay of $\phi$, entropy production during freeze-out may require a more careful treatment, as discussed in Sec.~\ref{sec:results} .
\item Finally, DM must be non-relativistic at freeze-out,
$T_{\rm f} < m_\chi/3.$ \cite{Kolb:1990vq}
\end{itemize}

When these conditions are satisfied, DM freeze-out occurs in a purely matter-dominated background, and the analytical treatment developed in the previous section provides an accurate description of the relic abundance.

It is important to note, however, that the parameter space beyond this region is not necessarily excluded, but instead requires a more careful analysis.

\section{Multipole Dark Matter: Theoretical Recipe}
\label{sec:multipole_theory}

In this section, we define the electromagnetic interactions of an electrically neutral DM candidate and delineate their implications for DM annihilation and scattering processes. As previously stated, such a DM species can have effective couplings with the photon through electromagnetic moments. Upto mass-dimension six, these electromagnetic interactions are most generally specified by the CP-even magnetic dipole moment (MDM), charge radius (CR) and anapole (AP)  operators, and the CP-odd electric dipole moment (EDM) operator. Assuming a spin-1/2 fermion ($\chi$) as the DM candidate, these operators are explicitly defined as
\beq \label{eq:EFT}
\begin{split}
 &\mathcal{L}^{(5)}_{\text{MDM}}
=
\frac{g}{\Lambda}\,
\bar{\chi}\,\sigma^{\mu\nu}\chi\,
F_{\mu\nu} \qquad \mbox{(Magnetic Dipole)} \\
&\mathcal{L}^{(5)}_{\text{EDM}}=\frac{i g}{\Lambda}\, 
\bar{\chi}\,\sigma^{\mu\nu}\gamma^{5}\chi\,
F_{\mu\nu} \quad \mbox{(Electric Dipole)} \\
 &\mathcal{L}^{(6)}_{\text{CR}}=\frac{g}{\Lambda^2}\,\bar{\chi}\,\gamma^\mu \chi\,\partial^\nu F_{\mu\nu} \quad \mbox{(Charge Radius)} \\
  &\mathcal{L}^{(6)}_{\text{AP}}= \mathcal{C}_{\rm AP}\frac{g}{\Lambda^2}\,\bar{\chi}\,\gamma^\mu\gamma^5 \chi\,\partial^\nu F_{\mu\nu} \quad \mbox{(Anapole)}
\end{split} \, ,
\eeq
with coefficients characterized by a dimensionless parameter $g$ and a cut-off scale $\Lambda$ \footnote{Presumably, these operators are loop-induced moments generated by a heavy, charged, dark sector particle running in the loop with a characteristic scale $\sim \Lambda$.}. Here, $F_{\mu \nu}$ is the electromagnetic field strength and $\sigma^{\mu \nu} = (i/2) [ \gamma^{\mu},\gamma^{\nu}]$ denotes the spin tensor. Although the preceding list of operators are most generally allowed for a Dirac fermion, it must be noted that for a Majorana fermion the dipole moments and the charge radius operator vanish identically, leaving the anapole moment as the leading gauge-invariant coupling to the photon ~\cite{Radescu:1985, Ho:2012bg}. Accounting for this possibility we include in $\mathcal{L}_{\text{AP}}$ the prefactor $\mathcal{C}_{\rm AP} = 1(1/2)$ for Dirac (Majorana) fermions.

As the processes governing the DM phenomenology are annihilation (to SM particles) and scattering (off nucleons), it is perhaps instructive to get a qualitative idea of the possible effects of the aforementioned higher-dimensional operators on the nature of the corresponding cross-sections. In particular, an estimate of the  energy-momentum dependence of the said cross-sections would be directly relevant in the discussions to follow regarding the DM's freeze-out dynamics as well as the constraints derived from direct detection experiments and indirect observations. To address these aspects systematically, we note that DM annihilations proceed either through an $s$-channel exchange of a photon or through a $t$-channel process mediated by the DM propagator itself. On the other hand, the DM-nucleon scattering, understandably, proceeds solely via a $t$-channel photon exchange. For brevity, we discuss the photon and DM mediated processes individually as follows--- 

\underline{\emph{Photon exchange}}:
The $2 \to 2$ $s$-channel processes in this case are $\chi \Bar{\chi} \to f\Bar{f}, \, W^{+} W^{-}$, where $f$ denotes the SM fermions. The $t$-channel exchange corresponds to the DM-nucleon scattering $\chi N \to \chi N$.
 The amplitudes of these processes can be written in a general current-current interaction form, namely
\beq
\mathcal{M} = \frac{J^{\mu}_{\chi}(p_1,p_2;q)J_{\mu, \rm SM}(k_1,k_2;q)}{q^2} \, ,
\eeq
where the factor $J^{\mu}_{\chi}$ specifies the dark sector current originating from the multipole operators and $J^{\mu}_{\rm SM}$ denotes the contribution emanating from a purely SM current. The currents in general depend on the momenta of the DM and the visible sector particles, namely $(p_1, p_2)$ and $(k_1, k_2)$, respectively. The variable $q$ denotes the momentum exchange flowing through the photon propagator, which goes as $q^2 = s$ for $s$-channel annihilations and $q^2 = t$ for scatterings, where $s$ and $t$ are the usual Mandelstam variables.
With the above definition in place, the following remarks are now in order---

\begin{itemize}
    \item Firstly, the derivatives in the operators have a contrasting, and in fact opposite, role to play in DM annihilations as compared to scattering processes. The said processes being mediated by a photon, the derivative couplings simply translate to an additional dependence on the momentum exchange ($q$) via the photon propagator. Therefore, the net dependence of the dark sector currents on $q$ can be readily factored out as
    
    \beq
    \begin{split}
        J_{\chi, \mu} &\propto (q_{\rho}\eta_{\mu \sigma})\mathcal{J}_{\chi}^{\rho \sigma} \qquad \quad (\rm for \, dipoles) \\
        &\propto (q^2\eta_{\mu \nu}-q_{\mu} q_{\nu})\mathcal{J}_{\chi}^{\nu} \quad (\rm for \, AP \, and \, CR)
    \end{split}
    \eeq
    where $\mathcal{J}_{\chi}$'s encapsulate the spinorial pieces of the currents. 
    
Clearly then, the explicit dependence on momentum transfer in the currents leads to an enhancement in the cross-sections in the case of DM annihilations (when $q^2 = s$) and to a suppression in the case of a low-energy scattering off a nucleon (when $q^2 = t$). By comparing the Lorentz structure of the currents, we see that the annihilation cross-section in the case of an anapole or charge radius DM is expected to receive a larger enhancement (by a factor of $s \sim 4 m_{\chi}^2$) as compared to that for a dipole DM. By the same argument, then, the scattering cross-section involving an AP or CR DM would be relatively more suppressed.

\item Apart from the dependence on the momentum exchange, the amplitudes also depend on the momenta of the incoming DM particle(s). The order at which the three-momenta of the DM particles enter the amplitudes determines the nature of the annihilation cross-sections vis-à-vis their dominant partial-wave behavior. The expected behavior can already be gleaned at the Lagrangian level, atleast qualitatively, by looking at the transformation properties of the operators under parity. With the P-even operators (MDM and CR), the DM annihilation cross-sections is understood to be dominantly $s$-wave in nature, whereas a $p$-wave nature is expected with an odd number of insertions of the P-odd ones (AP and EDM). This is most explicitly realized by reducing the spinor structure of the operators  in the nonrelativistic (NR) limit and keeping terms to lowest order in the DM three-momenta. For instance, for the specific case of an anapole DM which is a Dirac fermion annihilating to a pair of SM fermions ($f$), the amplitude in the CM frame in the NR limit reduces to
\beq
\begin{split}
    &\left[\bar{v}_{\chi}(p_2)\gamma^{\mu}\gamma^5 u_{\chi}(p_1)\right]\left[\bar{u}_{f}(k_1)\gamma_{\mu}v_{f}(k_2)\right] \\
    &\overset{\text{NR}}{\longrightarrow} 2 \, i \, m_f\left[\xi_2^{\dagger}(\vec{p}_1\times \vec{\sigma})_i\xi_1 \right]\left[ \phi_1^{\dagger}\sigma_i \phi_2\right] + \mathcal{O}\left(\vec{p}^{\, 3}\right)
\end{split}
\eeq
where $\sigma_i$ are the usual Pauli spin matrices and $\xi_{1,2}$ are the two-component Pauli spinors corresponding to $(u,v)_{\chi}$, respectively, and so are $\phi_{1,2}$ for the SM fermions. Being linear in $\vec{p}$ at the leading order, the NR amplitude clearly suggests a $p$-wave nature of the cross-section associated with it. For reference, a complete list of the NR limits and the corresponding partial-wave behavior of the multipole operators is summarized in Table \ref{tab:nrlimits}. In the following subsection, we show explicitly the velocity-dependence of thermally-averaged cross-section $\langle\sigma v_{\rm rel}\rangle$ for self-annihilation and scattering processes. We show, in particular, for the case of Anapole DM, that indeed $\langle\sigma v_{\rm rel}\rangle \propto v^2_{\rm rel}$.  
\end{itemize}

\underline{\emph{DM exchange}}: The remaining channel for DM annihilation, \emph{viz.} $\chi \Bar{\chi} \to \gamma \gamma$, is a $t$-channel process involving a DM propagator. Notably, the cross-section for the process vanishes identically in the case of anapole and charge radius operators. For insertions of dipole operators the general form of the amplitudes can be written as
\beq
\begin{split}
    \mathcal{M}_t \propto \frac{\Bar{v}_{\chi}(p_2)\Gamma^{\mu \nu}(\slashed{p}_1-\slashed{k}_1+m_{\chi})\Gamma^{\rho \sigma}u_{\chi}(p_1)}{t - m_{\chi}^2}&k_{2 \mu} k_{1 \rho}\\
    \times \epsilon^{*}_{\nu}&(k_2)\epsilon^{*}_{\sigma}(k_1)
\end{split}
\eeq
\beq
\begin{split}
    \mathcal{M}_u \propto \frac{\Bar{v}_{\chi}(p_2)\Gamma^{\mu \nu}(\slashed{p}_1-\slashed{k}_2+m_{\chi})\Gamma^{\rho \sigma}u_{\chi}(p_1)}{u - m_{\chi}^2}&k_{1 \mu} k_{2 \rho}\\
    \times \epsilon^{*}_{\nu}&(k_1)\epsilon^{*}_{\sigma}(k_2)
\end{split}
\eeq
for the $t$ and $u$-channels, respectively. Here, $k_{1,2}$ are the photon momenta and $\Gamma^{\mu \nu}$ acts as a placeholder for the gamma matrices present in the MDM and EDM operators.

Following the arguments made in the preceding segment, it is readily inferred that the cross-section for the diphoton process will always be dominantly $s$-wave in nature. This is true even for an EDM DM due to the presence of two insertions of the P-odd operator in the amplitude. Also, since in the NR regime $t=u\approx -m_{\chi}^2$ and $m_{\chi}$ is the only mass scale present in the process, it is easy to estimate from dimensional analysis that the cross-section would scale as $\sigma_{\gamma \gamma} \propto m_{\chi}^2/\Lambda^4$ for both MDM and EDM couplings.

Having illustrated the salient features of the various processes involving multipole DM operators, we present in the following subsection the expressions for the thermally-averaged annihilation cross-sections $\langle\sigma v_{\rm rel}\rangle_{\rm ann}$. We find that the explicit momentum dependence of $\langle\sigma v_{\rm rel}\rangle$ is in agreement with the preceding assertions. The detailed derivations of $\langle\sigma v_{\rm rel}\rangle$ are given in App.~\ref{sec:cross-section}. The expressions for the DM--nucleon scattering cross-sections for these operators can be found in~\cite{DelNobile:2021wmp, Kumar:2025wsp}.

\subsection{Expressions for Annihilation Cross-sections}\label{sec:X-section}

Here we make a list of the thermally-averaged annihilation cross-section formulae in the NR limit as a power series expansion in the relative velocity  $v_{\rm rel}$ of the incoming DM particles.\footnote{The thermally-averaged relative velocity depends on the bath temperature $T$ as $\langle v_{\rm rel}^2\rangle =6x^{-1}$, where, $x = m_\chi/T$.} We invoke for the relevant processes only one multipole DM operator at a time\footnote{While this might seem an \emph{ad hoc} assumption, it is quite possible to have DM models that dominantly induce only one of these operators in the IR. \emph{E.g.} the anapole operator is the only electromagnetic interaction possible for a Majorana DM as previously mentioned; for a Dirac fermion DM the specific CP nature of the UV physics can lead to the generation of either the EDM or the MDM operator at the leading order. Naively, it is only in the case of the CR operator that one needs to assume a fine tuned C and P-symmetric UV model so as to suppress the possible dim-5 MDM operator in the IR relative to the dim-6 CR operator. However, a discussion on the possible UV completions is beyond the scope of this work.} and categorize the expressions accordingly as follows---

\subsubsection*{Anapole}
\beq
~&\langle \sigma v_{\text{rel}} \rangle_{\chi \Bar{\chi} \to f\bar{f}}= \mathcal{C}_{\rm AP}^2\frac{2C_F\alpha g^2 m_\chi^2}{3\Lambda^4} \\
& \qquad \qquad  \qquad \times
\sqrt{1 - \frac{m_f^2}{m_\chi^2}}\left(1 + \frac{m_f^2}{2m_\chi^2}\right)\langle v_{\rm rel}^2 \rangle \, , \\
~&\langle \sigma v_{\text{rel}} \rangle_{\chi \Bar{\chi} \to W^+ W^-}=\frac{2\alpha g^2 m_\chi^6}{3\Lambda^4 m_W^4}\langle v_{\text{rel}}^2 \rangle \, ,
\eeq
%
%%%%%%%%%%%%%%%%%%%%%%%%%%%%  Dipole DM %%%%%%%%%%%%%%%%%%%%%%%%%%%%%%%%%%%%%%%%%%%%%%%%%%%%%%%%%%%

\subsubsection*{Electric Dipole Moment}
\beq
~&\langle \sigma v_{\text{rel}} \rangle_{\chi \Bar{\chi} \to f \bar{f}}=
\frac{C_F \alpha g^2 }{3\Lambda^2}\\
& \qquad \qquad \qquad \times \sqrt{1 - \frac{m_f^2}{m_\chi^2}}\left(1 + \frac{m_f^2}{2 m_\chi^2}\right)
\langle v_{\text{rel}}^2 \rangle,\\
~&\langle \sigma v_{\text{rel}} \rangle_{\chi \Bar{\chi}\to W^+ W^-}
=\frac{\alpha g^2 m_\chi^4}{3\Lambda^2 m_W^4}\langle v_{\text{rel}}^2 \rangle\, , \\
~&\langle \sigma v_{\rm rel} \rangle_{\chi \Bar{\chi}\to \gamma\gamma}=
\frac{g^4 m_\chi^2}{\pi \Lambda^4}
\left(8 + 5 \langle v_{\rm rel}^2 \rangle \right)\, ,
\eeq
\subsubsection*{Magnetic Dipole Moment}
\beq
~&\langle \sigma v_{\rm rel} \rangle_{\chi \chi \to f \bar{f}}=
\frac{4C_F\alpha g^2}{\Lambda^2}
\left(1 - \frac{1}{24}\langle v_{\rm rel}^2 \rangle \right),\\
~&\langle \sigma v_{\rm rel} \rangle_{\chi\chi\to W^+ W^-}
= \frac{4 \alpha g^2 m_\chi^4}{\Lambda^2 m_W^4}
\left(1 + \frac{\langle v_{\rm rel}^2\rangle}{3}\right),\\
~& \langle \sigma v_{\rm rel} \rangle_{\chi\chi\to \gamma \gamma}
=\frac{g^4 m_\chi^2}{\pi \Lambda^4}
\left(8 + 5 \langle v_{\rm rel}^2 \rangle\right),
\eeq
\subsubsection*{Charge Radius Operator}
\beq
~&\langle \sigma v_{\rm rel} \rangle_{\chi \chi \to f \bar{f}}=
\frac{C_F\alpha g^2 m_\chi^2}{\Lambda^4}
\left(4 + \frac{7}{6}\langle v_{\rm rel}^2 \rangle \right) ,\\
~& \langle \sigma v_{\rm rel} \rangle_{\chi\chi\to W^+W^-}=\frac{\,\alpha\, g^2\, m_\chi^6}
{\,\Lambda^4\, m_W^4}
\left(4 + \frac{19}{6}\langle v_{\rm rel}^2 \rangle \right) 
\eeq
where $\alpha$ is the electromagnetic fine-structure constant and $C_F$ denotes the QCD color factor with a value $C_F=3 (1)$ for quarks (leptons).
It is immediately apparent from the above expressions that the $s$-channel annihilations for the P-even operators are dominantly $s$-wave in nature while those involving the P-odd operators are $p$-wave suppressed, as was expected. The $t$-channel annihilation to $\gamma \gamma$ also, expectedly, exhibits an $s$-wave behavior for both MDM and EDM operators. Although the $t$-channel cross-sections are $1/\Lambda^4$ suppressed, the additional $m_{\chi}^2$ scaling of the same can render them comparable to the $s$-channel rates for large enough DM masses. For an EDM DM, the relative enhancement can be more pronounced due to the $p$-wave suppression of the $s$-channel processes. We further note that beyond the $W$-mass threshold, $m_{\chi}>m_W$, the $W^+ W^-$ channel dominates due to the apparent scaling of the cross-sections as $\sigma \sim m_{\chi}^4,m_{\chi}^6$ obtained for dim-5 and dim-6 operators, respectively. However, these scalings are spurious in nature and should be understood as artefacts of the limited validity of the EFT that we assume. We further elaborate on this point in the following section.

\subsection{A note on EFT validity}\label{sec:EFT-validity}

As pointed out in the preceding section, the $W^{+}W^{-}$ annihilation cross-section through a photon exhibits a spurious energy growth, \emph{viz.} $\sigma \sim s^2,s^3$ for dim-5 and dim-6 operators, respectively. While a pathological energy growth is indeed expected on account of the nonrenormalizable nature of the multipole operators, a naive dimensional analysis suggests that the cross-sections, at the worst, should scale as $\sigma \sim s^{D-5}$ for one insertion of an operator of mass-dimension $D$. For instance, the process involving  the anapole operator should have a scaling $\sigma \sim s$, \emph{i.e.} similar to what is obtained for the $f\Bar{f}$ annihilation. This implies that the enhanced energy growth obtained in the computations must be unphysical. The regime $\mchi > m_W$, therefore, warrants caution since a naive inclusion of the $W^+W^-$ channel can result in misleading identifications of viable regions in the dark sector parameter space\footnote{This can be gleaned \emph{e.g.} from ref.~\cite{Gao:2013vfa}, which includes the $W^+W^-$ annihilation channel mediated by a photon and retains the spurious divergence in the DM analysis.}.

The bad high-energy behavior can be simply attributed to the fact that the inclusion of the $W$ bosons as physical degrees of freedom breaks the manifest gauge-invariance of the theory since the operators in Eq.~(\ref{eq:EFT}) are written in terms of the electromagnetic field strength $F_{\mu\nu}$. Thus, the EFT description employed in Eq.~(\ref{eq:EFT}) should be understood as the low-energy limit of a fully gauge-invariant theory defined above the EW symmetry breaking scale. In the unbroken phase, the multipole interactions are minimally expressed in terms of the hypercharge field strength $B_{\mu\nu}$, ensuring invariance under the full $SU(2)_L \times U(1)_Y$ gauge symmetry. After EW symmetry breaking, these induce couplings of the DM to both the photon and the $Z$ boson. Understandably, then, DM annihilation to $W^+W^-$ must necessarily include the $Z$ mediated channel as well in order to restore gauge invariance in the computations. This indeed gets rid of the spurious energy scalings as shown in \cite{Arina:2020mxo, Choi:2024uva}.

For $m_{\chi}>m_W$, a gauge-invariant treatment also opens the $\chi\bar{\chi}\to Zh$ channel and, therefore, a more comprehensive analysis may require DM operators involving the SM Higgs as well\footnote{For example, if the multipole moments are generated by a (hypercharged) heavy scalar field, it will in general couple to the SM Higgs as well.}. Pursuing this, however, is beyond the scope of the present work and for the sake of simplicity we restrict our analysis to the regime $m_{\chi} < m_W$ where the operators defined in Eq.~(\ref{eq:EFT}) provide a sufficient and well-behaved description of the target phenomenology.

In summary, the point of this discussion is to draw the readers' attention to the fact that the validity of the assumed EFT is not merely defined by the violation of perturbative unitarity near the UV cutoff scale $\Lambda$ but also by the breakdown of gauge-invariance beyond the EW scale.

\begin{table*}[t]
\centering
\small
\renewcommand{\arraystretch}{1.3}
\setlength{\tabcolsep}{10pt}
\begin{tabular}{@{}l l l l l@{}}
\toprule
    \textbf{\shortstack{Interaction\\ \,}} & \textbf{\shortstack{Relativistic \\ Operator}} & \textbf{\shortstack{P,C \\Properties}} & \textbf{\shortstack{NR structure \\
    in $\mathcal{M}_{\rm ann}$}} &  \textbf{\shortstack{Partial-wave \\ scaling of $\langle \sigma v \rangle_{\rm ann}$}} \\
\midrule

MDM &
$\bar{\chi}\sigma^{\mu\nu}\chi\,F_{\mu\nu}$ &
P-even, C-even &
$\xi_2^{\dagger} \, \vec{\sigma}\cdot \vec{E}\, \xi_1$  &
\qquad $s$-wave \\

EDM &
$\bar{\chi}\sigma^{\mu\nu}\gamma^5\chi\,F_{\mu\nu}$ &
P-odd, C-even &
$\xi_2^{\dagger} \, \vec{p}\cdot \vec{E}\, \xi_1$ &
\qquad $p$-wave  \\

Anapole &
$\bar{\chi}\gamma^{\mu}\gamma^5\chi\,J_{\mu}$ &
P-odd, C-odd &
$\xi_2^{\dagger}\,(\vec{p}_1\times \vec{\sigma})\cdot \vec{J}\,\xi_1$ &
\qquad $p$-wave  \\

Charge Radius &
$\bar{\chi}\gamma^{\mu}\chi\,J_{\mu}$ &
P-even, C-even &
$\xi_2^{\dagger} \, \vec{\sigma}\cdot \vec{J}\, \xi_1$ &
\qquad $s$-wave \\
\bottomrule
\end{tabular}
\caption{Non-relativistic structure of the multipole operators which appear in the amplitudes ($\mathcal{M}_{\rm ann}$) of the DM annihilation processes at the leading order in DM momenta. The partial-wave behavior of the corresponding cross-sections is also listed for one insertion of the individual operators. Here, $(\xi_1,\xi_2)$ are the two-component Pauli spinors corresponding to the Dirac spinors $(u,v)$ of the DM field $\chi$. The vector $J$ denotes the SM electromagnetic current.}
 \label{tab:nrlimits}
\end{table*}

\section{Multipole Dark Matter: Phenomenology, Results, \& Discussions}
\label{pheno}

In the following, we first discuss the experimental constraints (direct detection and IceCube observations) employed in our analysis.~We then consider a spin-$1/2$ fermionic DM candidate and study one effective operator at a time (\emph{i.e.}, setting the coefficients of all other operators to zero). For each operator, we present the DM relic density contours obtained for both the matter-dominated and radiation-dominated freeze-out,  and subsequently examine the viability of the corresponding parameter space in light of the current direct detection constraints.

\subsection{Experimental constraints }

\subsubsection*{Direct detection limits}

For the direct detection limits, we extract the exclusion bounds for the anapole, dipole, and charge radius operators from~\cite{PandaX:2023luminance, Liang:2024ecw, Ibarra:2024mpq}. Across the considered DM mass range, the leading bounds come from the electron recoil, Migdal, and nuclear recoil searches---PandaX-4T electron recoil limits~\cite{Liang:2024ecw} dominate at the lowest masses, the DarkSide-50 Migdal limit~\cite{Ibarra:2024mpq} dominates at the GeV scale, followed by the PandaX-4T nuclear recoil limits~\cite{PandaX:2023luminance}, after which the LZ nuclear recoil limits~\cite{Ibarra:2024mpq} provide the strongest constraints. The LZ limits are determined up to $10^3 \, {\rm GeV}$ in~\cite{Ibarra:2024mpq} utilizing \texttt{WimPyDD}~\cite{Jeong:2021bpl}. For larger DM masses, the limits from different direct detection experiments are presented in~\cite{Kumar:2025wsp}. The combined direct detection exclusion limits considered in our analysis are shown as the blue curve for each operators in Figs.~\ref{fig:anapole-CR-DM},~\ref{fig:EDM-MDM}.

\subsubsection*{Solar DM capture and GeV neutrinos}

Another compelling direction to look for DM signatures is via the capture of DM particles inside celestial bodies due to DM scattering with nuclei and their annihilation into SM states, which would eventually give rise to neutrinos. These neutrinos can propagate through the stellar interior, escape the celestial body, and eventually reach the terrestrial detectors~\cite{IceCube:2016dgk, Leane:2017vag, Bose:2021yhz, IceCube:2021xzo, Maity:2023rez, Bose:2023yll, Bose:2024wsh, Krishna:2025ncv, IceCube:2025fcu, Nguyen:2025ygc, Nguyen:2026nhe, Nguyen:2026apa}. Null observations of GeV neutrinos from the Sun can put limits on the DM couplings. In this work, we extract the limits on the anapole and dipole operators from~\cite{Bose:2023yll}. For the charge radius operator, we derive the limit, and a brief discussion follows.

The capture rate of DM particles inside the Sun is given by~\cite{Garani:2017jcj, Busoni:2017mhe, Bose:2023yll}
\begin{equation}\label{eq:C_sun}
\begin{aligned}
    C_\odot = & \sum_k \left( \dfrac{\rho_0}{m_\chi} \right) \int_0^{R_\odot} 4 \pi r^2 \, dr \\
    & \times \int_0^{u_{\rm esc}} du_\chi \left( \dfrac{f_{v_\odot}(u_\chi)}{u_\chi} \right) w(r) \, \Omega_k^-(w),
\end{aligned}
\end{equation}
where $\rho_0$ and $u_\chi$ are the local DM density and ambient DM velocity, respectively. The velocity distribution profile in the rest frame of the Sun, $f_{v_\odot}(u_\chi)$, is assumed to follow a Maxwell-Boltzmann distribution\footnote{The deviation from this distribution and its impact on the capture rate are discussed in~\cite{Bose:2022ola}.}. In Eq.~\eqref{eq:C_sun}, the velocity $w(r)$ denotes the DM velocity at a distance $r$ from the center of the Sun, and the summation over $k$ runs over the different nuclei inside the Sun. The term $\Omega^-_k(w)$ encapsulates the capture probability of a DM particle with velocity $w(r)$ interacting with the $k$-th nucleus, and is given by~\cite{Geytenbeek:2016nfg}
\begin{equation}
    \Omega_k^-(w) = n_k(r) \, w(r) \, \int_{\frac{m_\chi u_\chi^2}{2}}^{\frac{m_\chi w(r)^2 \zeta_k}{2 \zeta_{k,+}^2}} dE_{\rm nr} \, \dfrac{d \sigma_k}{dE_{\rm nr}},
\end{equation}
where $n_k$ and $m_k$ are the density and mass of the $k$-th nucleus inside the Sun, with $\zeta = (m_\chi / m_k)$ and $\zeta_{k,+} = (\zeta_k + 1)/2$. The solar profile is extracted from~\cite{Vinyoles:2016djt}. The differential recoil rate for the charge radius operator is given by~\cite{DelNobile:2021wmp, Kumar:2025wsp}
\begin{equation}\label{eq:dsigma_dEnr_CR}
    \left( \dfrac{d \sigma_k}{dE_{\rm nr}} \right)_{\rm CR} = \dfrac{2 \, g^2 \, \alpha}{\Lambda^4} \dfrac{Z_k^2 \, m_k}{u_\chi^2} F^2(E_{\rm nr}) \, ,
\end{equation}
where $Z_k$ is the atomic number of the $k$-th nucleus, and $F(E_{\rm ER})$ is the nuclear form factor, assumed to be the Helm form factor~\cite{Helm:1956zz}. For the DM mass $m_\chi > 5 \, {\rm GeV}$, evaporation is negligible~\cite{Liang:2018cjn, Garani:2021feo}, and the number of DM particles inside the Sun is dictated by capture and annihilation. Under the equilibrium condition, the annihilation rate can be related to the capture rate as $\Gamma_{\rm ann} = C_\odot / 2$. The annihilation of DM particles inside the Sun can produce secondary neutrinos, and the differential neutrino flux reaching the Earth can be written as
\beq\label{eq:E2dNdE_nu}
    E_\nu^2 \dfrac{d \Phi_\nu}{dE_\nu} = \dfrac{\Gamma_{\rm ann}}{4 \pi D_\odot^2} \times \left( E_\nu^2 \dfrac{dN_\nu}{dE_\nu} \right),
\eeq
where $D_\odot$ is the distance between the Sun and the Earth and $dN_\nu/dE_\nu$ is the differential neutrino spectrum within the solar environment, generated using \texttt{$\chi$aro$\nu$}~\cite{Liu:2020ckq} for each of the SM final states and then summed over with the appropriate branching ratios. The neutrinos produced at the center of the Sun propagate through the solar interior, vacuum, the Earth, and its atmosphere, which is computed using \texttt{nuSQuIDS}~\cite{Arguelles:2021twb}, including neutrino oscillation, scattering, and tau regeneration.

After calculating the differential flux from Eq.~\eqref{eq:E2dNdE_nu}, we compute the differential event rate following~\cite{Maity:2023rez, Bose:2023yll}. In Ref.~\cite{IceCube:2016dgk}, IceCube and DeepCore published three years of data on muon track-like events as a function of the solar opening angle. Using the expected DM-induced neutrino flux from the solar direction together with the observed data, we perform a $\chi^2$-analysis to determine the limits on the DM couplings at the $95\%$ confidence level. Later, the IceCube Collaboration has also analyzed 6.75 years of DeepCore data from the solar direction and provided upper limits on the DM annihilation rate for different SM final states~\cite{IceCube:2021xzo}. We have recast those bounds in our scenario using the appropriate branching ratios and found that the $\tau^+ \tau^-$ channel provides the most stringent bounds. The envelope of the bounds obtained above is calculated for the charge radius operator and is shown as the green curve in Fig.~\ref{fig:anapole-CR-DM}. For the anapole and magnetic dipole operators, the limits are shown in Figs.~\ref{fig:anapole-CR-DM} and \ref{fig:EDM-MDM}, respectively. For the electric dipole operator, the annihilation rate into fermions is suppressed, making the limits much weaker than those from direct detection. Therefore, we do not show them in this work.

\subsection{DM freeze-out phenomenology and Results}\label{sec:results}
In this section, we determine the coupling strength required to reproduce the observed DM relic density within this alternative cosmological framework, and we assess the resulting parameter space in light of complementary constraints.
Having set the stage in Sec.~\ref{sec:MDFO} by describing the freeze-out mechanism during a radiation-dominated as well as a matter-dominated era, we use the DM self-annihilation cross-sections derived in Sec.~\ref{sec:X-section} and construct contours in the DM mass--coupling plane which satisfy the present day DM relic density~\cite{Planck:2018vyg}. We superimpose the experimental constraints over the relic contours to identify the allowed parameter space, and thereby, illustrate the viability of the aforementioned multipole DM models. 

We show that the parameter space obtained under the standard radiation-dominated freeze-out is already highly constrained for electromagnetic multipole DM, in agreement with previous studies~\cite{Ibarra:2024mpq,Bose:2023yll}. In contrast, we find that an early matter-dominated epoch substantially enlarges the viable parameter space through entropy dilution, allowing considerably weaker couplings while reproducing the observed relic abundance, which is the key result of our work. Among the four operators considered here, the electric dipole interaction remains strongly constrained even in the presence of an early matter-dominated era.

\begin{figure}
    \centering
    \includegraphics[width=0.9\linewidth]{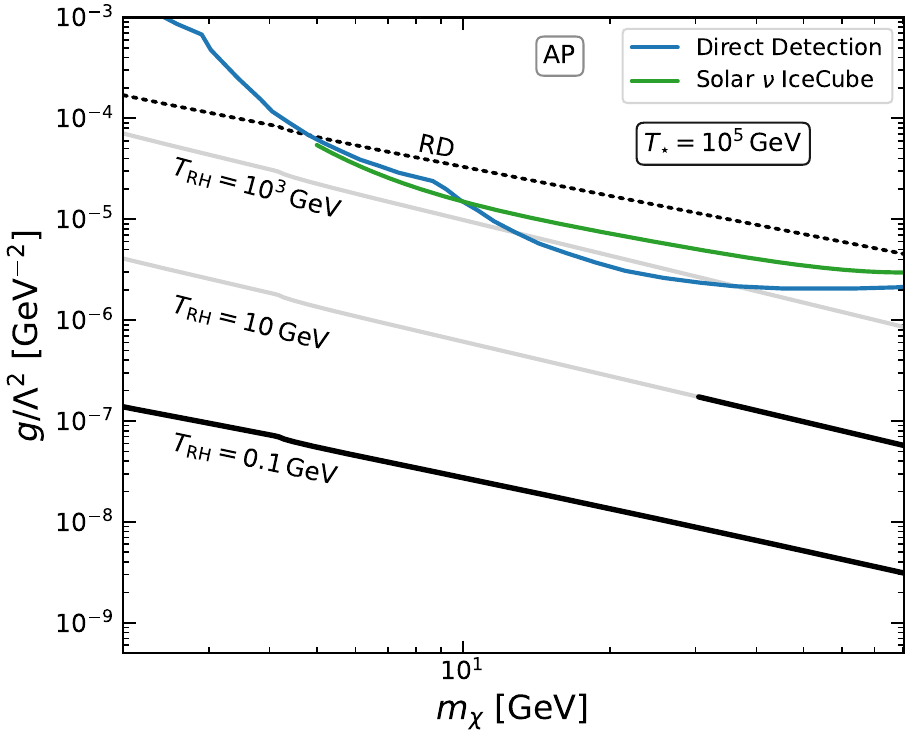}
    \includegraphics[width=0.9\linewidth]{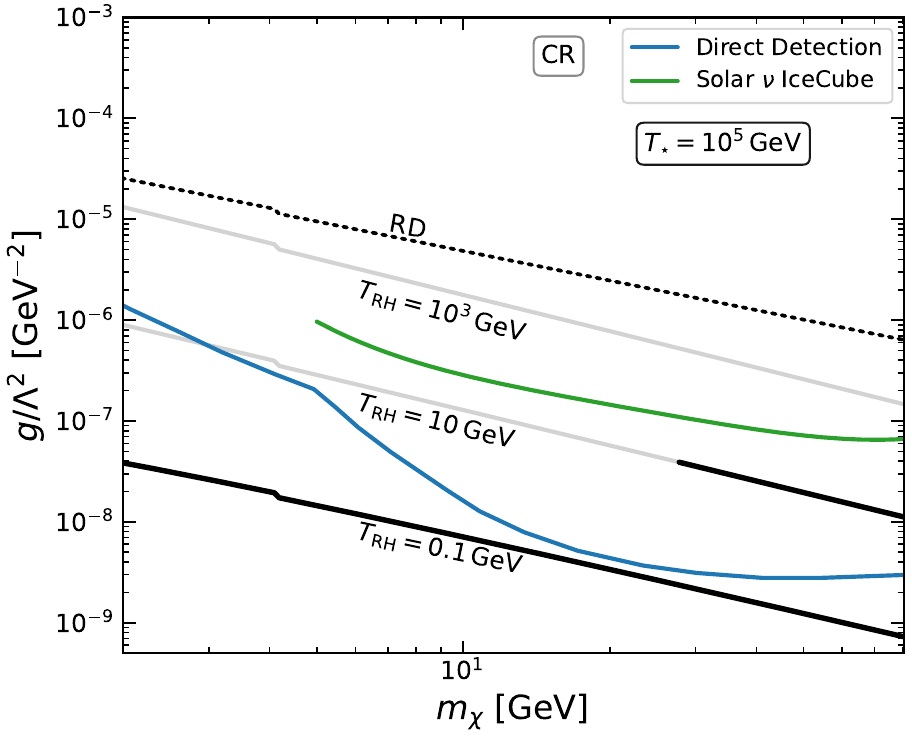}
    \caption{Dimension-six operators: relic density contours in the $(m_\chi,g/\Lambda^2)$ plane for the anapole interaction (top panel) and charge radius interaction (bottom panel), obtained by requiring the observed DM relic abundance, $\Omega_\chi h^2=0.12$, in the standard radiation-dominated (RD) cosmology and an early matter-dominated (MD) cosmology with varying reheating temperature $T_{\rm RH}$, and $r=0.99$. Current constraints from direct detection and solar neutrino searches are also shown. Along the MD relic density contours, the solid black segments correspond to $T_f>T_{\rm EV}$ (Eq.~\eqref{eq:T-EV}), while the gray segments correspond to $T_f<T_{\rm EV}$, where the freeze-out calculation requires a more careful treatment due to the significant decay of the matter field.}
\label{fig:anapole-CR-DM}
\end{figure}

\begin{figure}
    \centering
    \includegraphics[width=0.9\linewidth]{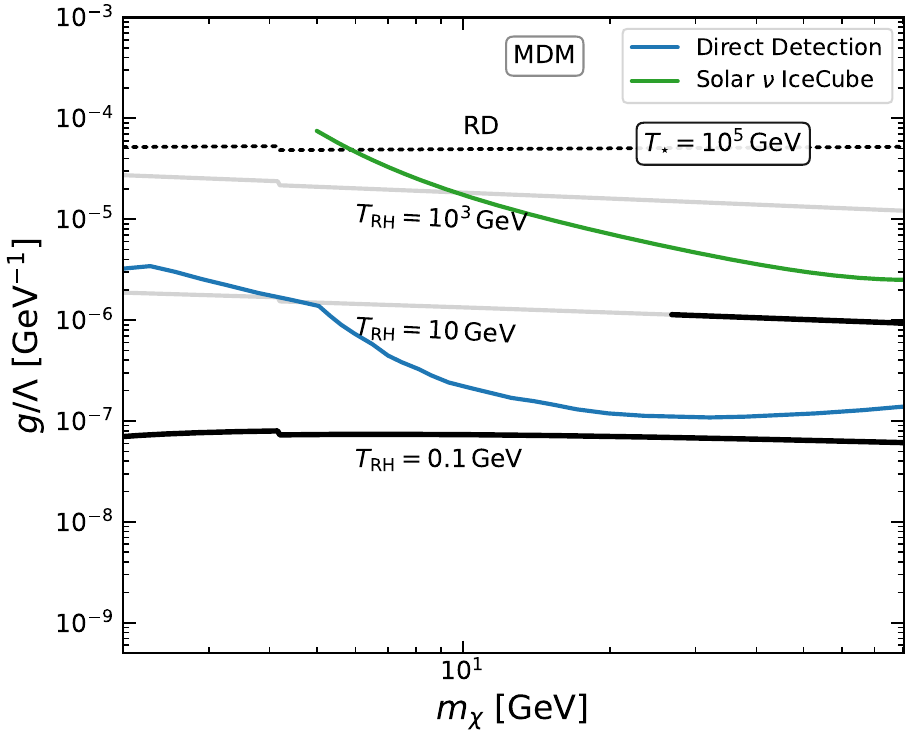}
    \includegraphics[width=0.9\linewidth]{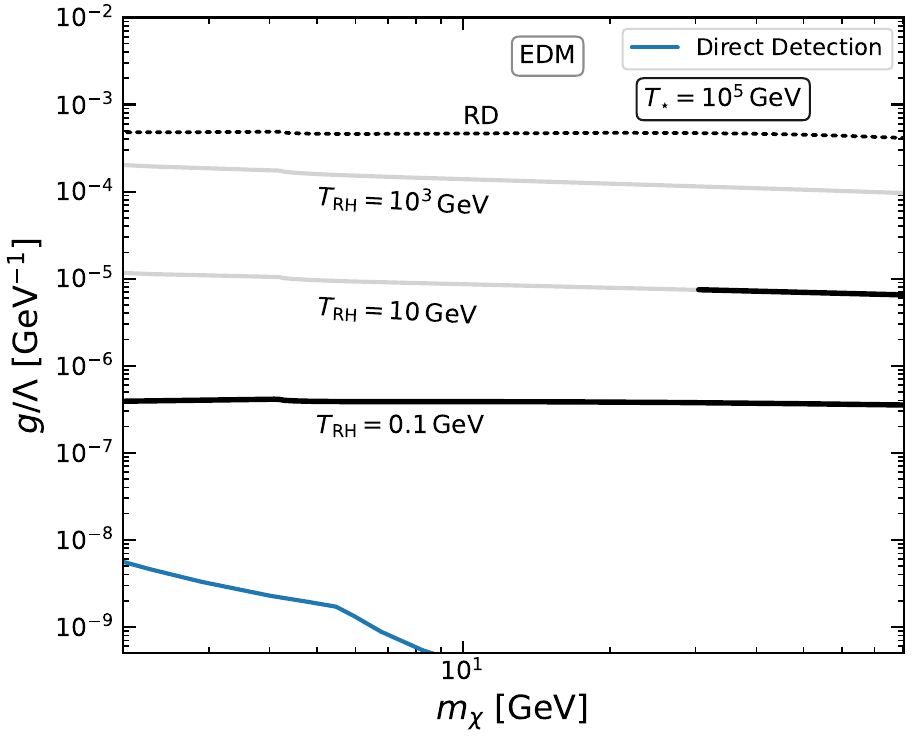}
 \caption{Dimension-five operators. relic density contours in the $(m_\chi,\, g/\Lambda)$ plane for the magnetic dipole interaction (top panel) and electric dipole interaction (bottom panel) operators, together with the corresponding constraints from direct detection constraints and solar neutrino searches. The contour definitions, and color codings are the same as in Fig.~\ref{fig:anapole-CR-DM}.} 
    \label{fig:EDM-MDM}
\end{figure}

Using the analytic expressions in Eqs.~\eqref{eq:MDFO-density}--\eqref{eq:xf-MD-RD}, we determine the couplings required to reproduce the observed DM relic abundance for each electromagnetic multipole operator. The resulting relic density contours for the anapole, charge radius, electric dipole, and magnetic dipole interactions are shown in Figs.~\ref{fig:anapole-CR-DM} and \ref{fig:EDM-MDM}, together with the envelopes of the current direct detection and IceCube constraints from Sun. Throughout this analysis we fix $T_\star=10^5~{\rm GeV},~ r = 0.99$, and consider three representative reheating temperatures, $T_{\rm RH}=0.1,10,$ and $10^3~{\rm GeV}$, characterizing the different amounts of entropy dilution while remaining safely above the BBN temperature.

As expected, the interaction strength required to obtain the observed relic abundance decreases as the reheating temperature is lowered. This effect can be understood directly from eqns.~\eqref{eq:zeta-TRH} and \eqref{eq:Tgamma-TRH} as the DM abundance dilution by entropy injection is correlated with the reheat temperature.  Note that the matter-dominated freeze-out depends on an extra energy scale $T_\star$, and, therefore, has a different relation to $m_\chi$ compared to the radiation-dominated freeze-out as can be seen from Eqs.~\eqref{eq:MDFO-density}--\eqref{eq:xf-MD-RD}. This results in the apparent morphological difference between the corresponding relic contours.

The morphology of the relic density contours can be understood by dividing the parameter space into two physically distinct regimes. The first corresponds to the region where annihilation is dominated by fermionic final states for DM mass $m_\chi<m_W$. For the dimension-six anapole and charge radius operators, the only relevant channel in this region is $\chi\Bar{\chi}\rightarrow f\bar f,$
since these operators do not induce the $\chi\Bar{\chi}\rightarrow\gamma\gamma$ final state at leading order. Consequently, the corresponding relic density contours vary smoothly below the $m_W$ threshold.

The second regime also lies below the $m_W$ threshold but is specific to the magnetic and electric dipole operators, which receive contributions from both
$\chi\bar{\chi}\rightarrow f\bar f$ and $\chi\bar{\chi}\rightarrow\gamma\gamma.$
The fermionic contribution scales as $\langle\sigma v\rangle_{\chi\Bar{\chi}\rightarrow f\bar f}\propto
\left(g/\Lambda\right)^2,$ whereas the diphoton channel scales as
$\langle\sigma v\rangle_{\chi\Bar{\chi}\rightarrow\gamma\gamma}\propto \left(g/\Lambda\right)^4m_\chi^2.$
At low reheating temperatures, the entropy dilution allows the observed relic abundance to be reproduced with substantially smaller effective couplings. The diphoton contribution is, therefore, suppressed relative to the fermionic channel which dominates the total annihilation cross-section. This explains why the $T_{\rm RH}=0.1~{\rm GeV}$ contours for the magnetic and electric dipole operators remain nearly flat for $m_\chi\ll m_W$. As the reheating temperature increases, the larger coupling required to obtain the correct relic abundance enhances the relative importance of the diphoton channel, producing the mild upward curvature observed below the electroweak threshold.

The phenomenological implications differ substantially among the four electromagnetic operators. For the anapole interaction, the relic density contours corresponding to $T_{\rm RH}=10~{\rm GeV}$ and $0.1~{\rm GeV}$ remain compatible with current direct detection and IceCube constraints over a substantial range of DM masses. For the magnetic dipole interaction and charge radius interaction, only the $T_{\rm RH}=0.1~{\rm GeV}$ contour survives the existing limits, whereas the contours corresponding to larger reheating temperatures are excluded. In contrast, the electric dipole interaction is considerably more constrained, with almost the entire parameter space explored in this work being excluded irrespective of the cosmological history.

The stronger direct detection sensitivity of the electric dipole operator originates from its non-relativistic interaction with the target nuclei. Unlike the magnetic dipole operator, the electric dipole interaction generates a coherent spin-independent contribution in addition to spin-dependent interactions~\cite{Anand:2013yka}, resulting in substantially stronger direct detection limits. Similar considerations apply to the charge radius operator, whose coherent interaction with nuclei also leads to stringent constraints from direct detection and GeV scale neutrino measurements from Sun at IceCube. Nevertheless, the entropy dilution associated with an early matter-dominated epoch enlarges the phenomenologically viable parameter space relative to the standard radiation-dominated freeze-out scenario. 

%Finally, we indicate in Figs.~\ref{fig:anapole-CR-DM} and \ref{fig:EDM-MDM} the regions satisfying the cosmological consistency conditions discussed in Sec.~\ref{sec:consistency1}. Since $T_\Gamma$ and $T_{\rm RH}$ are of the same order and both characterize the onset of significant radiation production from the decay of the heavy field $\phi$, we use the condition $T_f>T_\Gamma$ to identify the region where the freeze-out treatment adopted here is self-consistent. For the benchmark reheating temperatures considered, this condition mainly affects the $T_{\rm RH}=10^3~{\rm GeV}$ contours, whereas the $T_{\rm RH}=10~{\rm GeV}$ and $0.1~{\rm GeV}$ contours satisfy $T_f>T_\Gamma$ throughout the parameter space shown. The gray segments correspond to $T_f<T_\Gamma$, where DM freezes out after the onset of significant $\phi$ decay. These regions should not be regarded as excluded; rather, they indicate where a more complete treatment, including the coupled evolution of DM freeze-out, heavy-field decay, and entropy production, would be required. Such an analysis lies beyond the scope of the present work but is not expected to modify the central conclusion that an early matter-dominated epoch substantially enlarges the viable parameter space of electromagnetic multipole DM compared with the standard radiation-dominated cosmology.

%\subsubsection{Sudden Decay Approximation}\label{sec:sudden-decay}

The DM relic abundance calculation presented in Sec.~\ref{sec:DM-relic density} assumes entropy conservation until when $\phi$ decays instantaneously at a time $\sim \Gamma_\phi^{-1}$. However, a more natural expectation is that $\phi$ decays through an exponential law~\cite{Scherrer:1985}. In that case, while solving the Boltzmann equation the entropy injection in the thermal bath will be non-negligible and, therefore, the freeze-out calculation should be carried out in the line of Giudice-Kolb-Riotto~\cite{Giudice:2000ex,Pallis:2004yy}.

 Thus, the total radiation energy density is the old radiation in the thermal bath plus the new radiation $\rho_{R,\rm Decay}$  produced by $\phi$ decay
 \beq\label{eq:EV-rhoR}
 \rho_{R,\rm tot} = \rho_{R,\star}\left(\frac{a_{\star}}{a}\right)^4 + \rho_{R,\rm Decay}.
 \eeq
Scherrer and Turner~\cite{Scherrer:1985} pointed out that entropy of the bath remains nearly unchanged until the new radiation becomes comparable to the old radiation. The entropy violation time-frame can be derived by equating the first and second term of Eq.~\eqref{eq:EV-rhoR}, and, as shown in~\cite{Chanda:2019xyl} the temperature corresponding to the entropy-violation is given as
\beq\label{eq:T-EV}
T_{\rm EV}=T_\star\left(1+\frac{r}{1-r}\frac{v}{\tau_\star}\right)^{(1-v)/v},
\eeq
where, $v$ is defined as
\beq
v = \frac{2}{3}(1+w)^{-1}+1,
\eeq
with $w$ being the cosmological equation of state parameter relating pressure $p$ and energy density $\rho$ as $p = w \rho$. The dimensionless quantity $\tau = \Gamma_\phi t$.

For a continuous decay of $\phi$, the entropy of the radiation bath remains approximately conserved for $T_{\rm f}>T_{\rm EV}$. For $T_{\rm f}\lesssim T_{\rm EV}$, radiation produced by the decay of $\phi$ becomes important, and the evolution enters the decay-dominated regime, where $T\propto a^{-3/8}$ \cite{Giudice:2000ex}. The condition $T_f>T_\Gamma$ identifies freeze-out before
the instantaneous decay event in our sudden-decay approximation.
Continuous decay introduces a distinct scale, $T_{\rm EV}$, at which the decay-induced radiation becomes comparable to the pre-existing bath. In Figs.~\ref{fig:anapole-CR-DM} and \ref{fig:EDM-MDM}, we use $T_{\rm EV}$ as the consistency condition, as this is stronger by an order of magnitude\footnote{For example, with $T_{\rm RH}=10~{\rm GeV}$, $T_\star=10^5~{\rm GeV}$, and $r=0.99$, we estimate $T_\Gamma\simeq0.3~{\rm GeV}$ and $T_{\rm EV}\simeq2.6~{\rm GeV}$.} compared to $T_\Gamma$.

We also note that while the assumption of the electromagnetic-only DM couplings is consistent in its own accord as an EFT description below the EW scale, it disregards the possible effects of invoking $Z$-boson couplings for sub-EW scale DM masses. As shown in ref.~\cite{Arina:2020mxo}, these effects include the opening of the DM annihilation channel to $\nu \Bar{\nu}$ and the resonance enhancement of the individual annihilation cross-sections around the $Z$-pole. Thus, in connection to the relic-density calculations, this will invariably result in a dip in the associated contours in the coupling-vs-$m_{\chi}$ plane, a feature missing in our results in Figs.~\ref{fig:anapole-CR-DM} and \ref{fig:EDM-MDM}. While these deviations are quantitatively significant, their qualitative effect is simply identified as that it weakens the experimental and observational bounds on the parameter space around the resonance region, \emph{viz.} $m_{\chi} \sim m_{Z}/2$. Therefore, the expected changes will not invalidate any of our qualitative conclusions made within the purview of this work, which was to show how a nonstandard cosmological history can evade existing constraints on electromagnetically interacting DM. On the other hand, the aforementioned enhancement in the annihilation cross-sections may also affect the indirect detection limits, namely those derived from annihilations in galactic halos as well as from signals due to possible DM capture within celestial bodies. For the former, Ref.~\cite{Arina:2020mxo} (which assumes hypercharge couplings) finds that the bounds are typically weaker than those from direct detection; although they may probe parts of the radiation-dominated relic-density contours, they are not expected to exclude the lower-coupling contours for the matter-dominated scenarios. A similar conclusion holds true for the latter as well. This can be inferred by comparing the limits derived in this work using IceCube observations of solar neutrinos with those from a recent study \cite{Ibarra:2026otc} which has extended such capture-based probes to Earth, Jupiter, and white dwarfs.

\section{Conclusions}\label{sec:conclusions}
Electromagnetic multipole interactions provide a natural and minimal framework for an electrically neutral DM to couple to photons. 
In this work, we have shown that such interactions, although strongly constrained in the standard radiation-dominated freeze-out scenario, can remain phenomenologically viable in the presence of an early matter-dominated epoch. We have considered four multipole operators upto dimension-six, \emph{viz.} magnetic dipole moment, electric dipole moment, anapole, and charge-radius operators. For each case, we have determined the coupling required to reproduce the observed DM relic abundance in both radiation- and matter-dominated cosmologies, and confronted the resulting parameter space with current direct detection and IceCube solar neutrino constraints.

A generic feature of freeze-out during an early matter-dominated epoch is the entropy injection associated with the decay of the heavy matter field, $\phi$. The resulting dilution of the DM abundance reduces the interaction strength required to reproduce the observed relic density compared with the standard radiation-dominated scenario. Consequently, regions of parameter space that are strongly constrained, or even excluded, by direct detection experiments and IceCube observations of neutrinos from the Sun in the conventional cosmological picture become viable in a non-standard early cosmological history, which constitutes the key finding of this work. This effect is most pronounced for the anapole interaction, while the electric dipole operator remains strongly constrained over nearly the entire parameter space considered here. We have also identified the region where DM freeze-out occurs after the onset of significant decay of the heavy field. This region is not excluded, but instead indicates where the simple freeze-out treatment adopted here should be refined by consistently incorporating heavy-field decay and entropy production during freeze-out. Such refinements are not expected to alter the main conclusions of this work.

Several directions remain to be explored. For DM masses above the electroweak scale, additional annihilation channels become relevant. Although the inclusion of these channels would be an improvement, a more comprehensive treatment ultimately requires a complete gauge-invariant operator basis including DM-Higgs interactions \cite{Criado:2021trs}. We defer a dedicated investigation for the heavy-mass regime to a future work, where we address these aspects in detail for freeze-out as well as for the freeze-in production during the decay-dominated phase.

Our results highlight that non-standard cosmological histories can qualitatively alter the phenomenology of electromagnetic multipole DM, reopening regions of parameter space that appear excluded under the conventional radiation-dominated freeze-out paradigm. 

%
%
%%%%%%%%%%%%%%%%%%%%%%%%%%%%%%%%%%%%%%%%%%%%%%%%%%%%%%%%%%%%%%%%%%%%%%%%%%%
\vspace*{0.4in}
\emph{Note Added:} While we were finalizing this work, Ref.~\cite{Jahedi:2026mxa} appeared on arXiv, presenting a related study of dipole DM models in the context of non-standard cosmology. However, our work complements their analysis by considering all possible electromagnetic operators for neutral DM up to dimension six. Moreover, we point out the possible concerns and issues regarding the complete gauge invariance and EFT validity for DM masses beyond the EW scale, in contrast to the aforementioned paper.

%%%%%%%%%%%%%%%%%%%%%%%%%%%%%%%%%%%%%%%%%%%%%%%%%%%%%%%%%%%%%%%%%%%%%%%%%%%

%
%
%%%%%%%%%%%%%%%%%%%%%%%%%%%%%%%%%%%%%%%%%%%%%%%%%%%%%%%%%%%%%%%%%%%
\section*{Acknowledgments}

We sincerely thank Subhadip Bouri, Basudeb Dasgupta, Amol Dighe,  Biprajit Mondal, Rohan Pramanick, Akash Kumar Saha for their helpful discussions and insightful suggestions. We
are indebted to Debajyoti Choudhury, Ranjan Laha and James A. Unwin for useful discussions and for their valuable comments on the
manuscript. DB acknowledges the Council of Scientific and Industrial Research (CSIR), Government of India, for supporting his research under the Research Associateship program through grant no.\,\,09/0079(24106)/2025-EMR-I. PM, SM and PC acknowledge the support from the
Department of Atomic Energy (DAE), Government of India, under Project Identification
Number RTI 4002.
%%%%%%%%%%%%%%%%%%%%%%%%%%%%%%%%%%%%%%%%%%%%%%%%%%%%%%%%%%%%%%%%%%%
%
%

\appendix
\section{Solution of the Boltzmann equation}\label{sec:derivation-FO}
We start by defining a quantity $\Delta_\chi$ which is a measure of departure from the equilibrium abundance
\beq
\Delta_\chi = Y_\chi - Y_{\chi,\rm eq},
\eeq
the Boltzmann equation in Eq.~\eqref{eq:BE-eqn} can be rewritten as
\beq\label{eq:MDFO-A2}
\frac{d\Delta_\chi}{dx}= - \frac{dY_{\chi, \rm eq}}{dx}- \sum_n \lambda_n x^{-n-2} g_{\rm eff}^{1/2}
\left(\Delta_\chi^2 + 2 \Delta_\chi Y_{\chi,{\rm eq}}\right).
\eeq
At sufficiently late times, well after freeze-out ($x \gg x_f$), the temperature becomes low enough that $Y_\chi$ no longer follows its equilibrium value. In this regime the abundance $Y_\chi$ becomes comparable to $\Delta_\chi$, since $Y_{\chi,\rm eq}$ is exponentially suppressed. Consequently, terms in Eq.~\eqref{eq:MDFO-A2} involving $dY_{\chi,\rm eq}/dx$ and $Y_{\chi,\rm eq}$ can be neglected, and the evolution equation simplifies to
\beq
\frac{d\Delta_\chi}{dx}\simeq- \sum_n \lambda_n g_{\rm eff}^{1/2} x^{-n-2}\Delta_\chi^{2}.
\eeq
Integrating this expression over the interval $[x_f,\infty)$ yields
\beq
\int_{x_{\rm f}}^{\infty}\frac{d\Delta_\chi}{\Delta_\chi^{2}}\simeq
- \int_{x_{\rm f}}^{\infty}dx \sum_n \lambda_n g_{\rm eff}^{1/2} x^{-n-2}.
\eeq

Evaluating the left-hand side gives
\beq
\frac{1}{\Delta_{\chi,\infty}}=\frac{1}{\Delta_{\chi,{\rm f}}}+\sum_n \int_{x_{\rm f}}^{\infty}dx\, \lambda_n g_{\rm eff}^{1/2} x^{-n-2}.
\eeq
Usually the second term in the RHS dominates.

It follows that the asymptotic anti-dark-matter abundance is
\beq
Y_{\chi,\infty}\simeq\Delta_{\chi,\infty}=\frac{1}{\sum_n \int_{x_{\rm f}}^{\infty}dx\, \lambda_n g_{\rm eff}^{1/2} x^{-n-2}}.
\eeq
Thus, substituting for $\lambda_n$ and $g_{\rm eff}^{1/2}$
\beq
Y_{\chi,\infty}\simeq \frac{1}{\sqrt{\frac{\pi}{45}}m_\chi M_{\rm Pl}g_\ast^{1/2}(1-r)^{-1/2}\sum_{n}\frac{\sigma_n x_{\rm f}^{-(n+3/2)}}{{n+3/2}}x_\star^{1/2}},
\eeq
where,~we~have observed that for our choice of parameters, $g_{\rm eff}^{1/2}$ $\,\simeq\,(3/4)g_{\star}^{1/2}(1-r)^{-1/2}x^{-1/2}x_{\star}^{1/2}$.
Therefore, including the dilution factor the DM relic density is 
\beq
\Omega_{\rm DM}^{\rm Relic} h^2 &= \zeta\frac{s_0 m_\chi}{\rho_c}h^2Y_{\chi, \infty}\\
& \simeq \zeta\frac{1.0\times 10^9  ~{\rm GeV}^{-1}}{M_{\rm Pl} g_\star^{1/2}(1-r)^{-1/2}}\left(\sum_{n}\frac{\sigma_n x_{\rm f}^{-(n+3/2)}}{{n+3/2}}x_\star^{1/2}\right)^{-1},
\eeq
where $\rho_c = 3H_0^2M_{\rm Pl}^2/(8\pi)$, and $s_0 \simeq 2970~{\rm cm}^{-3}$.
%%%%%%%%%%%%%%%%%%%%%%%%%%%%%%%%%%%%%%%%%%%%%%%%%%%%%%%
\subsection{Calculation for the freeze-out temperature}
%%%%%%%%%%%%%%%%%%%%%%%%%%%%%%%%%%%%%%%%%%%%%%%%%%%%%%%
At freeze-out the DM relic abundance stop tracking the equilibrium abundance, hence,
\beq
\Delta_\chi(x_{\rm f}) \simeq Y_{\chi, \rm eq}(x_{\rm f}) = \delta_\chi Y_{\chi,\rm eq}(x_{\rm f}),
\eeq
where $\delta_\chi$ is a order of unity factor.

Neglecting $d\Delta_\chi/dx$ near freeze-out, from Eq.~\eqref{eq:MDFO-A2} we may write
\beq
\left.\frac{dY_{\rm\chi}}{dx}\right|_{\rm f}
\approx
-\sum_n \lambda_n\,\delta_\chi(\delta_\chi+2)\,x_{\rm f}^{-n-2}
\,g_{\rm eff}^{1/2}(x_{\rm f})Y_{\chi,{\rm eq}}^{2}(x_{\rm f}).
\eeq
%
% where,
% %
% \beq
% Y_{\chi}^2 - Y_{\chi, \rm eq}^2 = 2Y_{\chi,\rm eq}\Delta_\chi +\Delta_\chi^2 = Y_{\chi,\rm eq}^2(2\delta_\chi+\delta_\chi^2).
% \eeq
% %

Also, 
\beq
\frac{dY_{\chi,\rm eq}}{dx} = Y_{\chi,\rm eq}\left(\frac{3}{2x}-1\right),
\eeq
thus using the approximation $\left.dY_{\rm\chi,eq}/dx\right|_{\rm FO}\approx\left.dY_\chi/dx\right|_{\rm FO}$ we obtain
\beq
\left(\frac{3}{2x_{\rm f}}-1\right) \approx - \left[\sum_n\delta_\chi(\delta_\chi+2)\lambda_n x_{\rm f}^{-n-2}g_{\rm eff}^{1/2}\right]Y_{\chi,{\rm eq}}(x_{\rm f}).
\eeq

Substituting for the equilibrium abundance $Y_{\chi,{\rm eq}}$ from Eq.~\eqref{eq:Yeq} and using $\left(1-\frac{3}{2x_{\rm f}}\right)\approx 1$, we obtain
\beq
x_{\rm f} = \ln\left(b\left[\sum_n\delta_\chi(\delta_\chi+2)\lambda_n x_{\rm f}^{-n-1/2}g_{\rm eff}^{1/2}\right]\right).
\eeq
As argued by Scherrer \& Turner~\cite{Scherrer:1985zt}, the numerical results suggest a best fit $(2\delta_\chi+\delta_\chi^2)\simeq n+1$.
Substituting for $g_{\rm eff}$, $\lambda_n$, and $b$, we obtain
\beq
x_{\rm f}^{\rm MD}& = \ln\left[\frac{\sqrt{45}g_\chi}{\sqrt{32}\pi^3}g_{\star}^{-1/2}\frac{M_{\rm Pl}m_{\chi}^{3/2}}{(1-r)^{1/2}T_{\star}^{1/2}}\sum_n\frac{(n+1)\sigma_{n}}{x_{\rm f}^{n+1}}\right],\\
x_{\rm f}^{\rm RD}& = \ln\left[\frac{\sqrt{45}g_\chi}{\sqrt{32}\pi^3} g_{\star}^{-1/2}M_{\rm Pl}m_{\chi}\sum_n\frac{(n+1)\sigma_{n}}{x_{\rm f}^{n+1/2}}\right].
\eeq
\section{Thermally averaged annihilation cross-section calculation}\label{sec:cross-section}

\subsubsection{ Anapole Moment (AP)}
% If kinematically allowed, $\chi$ can annihilate into all the SM leptons and quarks as well as into the $W$ boson. 
The differential cross-section for the DM annihilation to the SM fermions ($f$) in the center-of-mass (CM) frame is given by
\begin{equation}
\frac{d\sigma}{d\Omega}\bigg|_{\chi\Bar{\chi}\to f\bar{f}}=\frac{1}{64\pi^2 s}
\sqrt{\frac{s - 4 m_f^2}{s - 4 m_\chi^2}}
    \,|\overline{\mathcal{M}}_{f \bar{f}}|^2 ,
\end{equation}
where the spin-averaged squared amplitude reads
\beq
|\overline{\mathcal{M}}_{f \bar{f}}|^2&= \mathcal{C}_{\rm AP}^2\frac{2 e^2 g^2 C_F}{\Lambda^4}
\Big[2 m_\chi^4- 4 m_\chi^2 (m_f^2 + s + t) \\
& \hspace{1.0cm} + 2 m_f^4 - 4 m_f^2 t + s^2+ 2 s t+ 2 t^2\Big]\,,
\eeq
in terms of Mandelstam variables $s$, $t$ and $u$, and the SM fermion masses $m_f$. In the above, $C_F$ denotes the color factor with a value $C_F=3 (1)$ for quarks (leptons), and the parameter $\mathcal{C}_{\rm AP}$ specifies whether the DM is a Dirac or Majorana fermion with the entry $\mathcal{C}_{\rm AP}=1(1/2)$ for Dirac (Majorana) fermion.

In the NR limit the thermally-averaged annihilation cross-section for the aforementioned process can be obtained as a power series expansion in the relative velocity of the incoming DM particles $v_{\rm rel}$. At the leading order in $v_{rel}$, this can be expressed as
%
% \beq
% \langle \sigma v_{\text{rel}} \rangle_{\chi\chi \to f\bar{f}}
% &=
% \frac{\alpha_{\rm em} g^2 m_\chi^2}{8\Lambda^4}
% \sqrt{1 - \frac{m_f^2}{m_\chi^2}}
% \\
% &\hspace{-0.6cm}\times
% \Bigg[
% \frac{4}{3}\left(1 - \frac{m_f^2}{m_\chi^2}\right)
% + 4\left(1 + \frac{m_f^2}{m_\chi^2}\right)
% \Bigg]\langle v_{\rm rel}^2 \rangle \, .
% \eeq
%
\beq
~&\langle \sigma v_{\text{rel}} \rangle_{\chi \Bar{\chi} \to f\bar{f}}
= 
\mathcal{C}_{\rm AP}^2\frac{2C_F\alpha g^2 m_\chi^2}{3\Lambda^4} \\
& \qquad \qquad  \qquad \times
\sqrt{1 - \frac{m_f^2}{m_\chi^2}}\left(1 + \frac{m_f^2}{2m_\chi^2}\right)\langle v_{\rm rel}^2 \rangle \, ,
\eeq

Similarly, the
differential cross-section for the DM annihilation to $W^+ W^-$ in the CM frame is
given by,
\begin{equation}
\frac{d\sigma}{d\Omega}\bigg|_{\chi\Bar{\chi}\to W^+ W^-}=\frac{1}{64\pi^2 s}
\sqrt{\frac{s - 4 m_W^2}{s - 4 m_\chi^2}}
\,|\overline{\mathcal{M}}_{W^+ W^-}|^2 ,
\end{equation}

% where
% \beq
% |\overline{\mathcal{M}}_{W^+ W^-}|^2&= 
% -\frac{e^2 g^2}{2 \Lambda^4 m_W^4}
% \Big[
% m_\chi^4 \left(12 m_W^4 - 4 m_W^2 s + s^2 \right) 
% \\
% & \hspace{1.0 cm}
% -2 m_\chi^2 (12 m_W^6 + 4 m_W^4 (7 s + 3 t) 
% \\
% & \hspace{1.0 cm}
% - m_W^2 s (7 s + 4 t) + s^2 t )+12 m_W^8
% \\
% & \hspace{1.0 cm}
% -4 m_W^6 (s + 6 t)
% +m_W^4 (17 s^2 + 20 s t + 12 t^2)
% \\
% & \hspace{1.0 cm}
% -2 m_W^2 s (2 s^2 + 3 s t + 2 t^2)
% +s^2 t (s + t)
% \Big] \, .
% \eeq

\begin{align}
\label{eq:anapoleWW}
|\overline{\mathcal{M}}_{W^+ W^-}|^2
&= -\frac{e^2 g^2}{2 \Lambda^4 m_W^4}
\Big[
m_\chi^4 (12 m_W^4 - 4 m_W^2 s + s^2)
\nonumber \\
&\quad - 2 m_\chi^2 \Big(
12 m_W^6 + 4 m_W^4 (7 s + 3 t)
\nonumber \\
&\qquad - m_W^2 s (7 s + 4 t) + s^2 t
\Big)
\nonumber \\
&\quad + 12 m_W^8 - 4 m_W^6 (s + 6 t)
\nonumber \\
&\quad + m_W^4 (17 s^2 + 20 s t + 12 t^2)
\nonumber \\
&\quad - 2 m_W^2 s (2 s^2 + 3 s t + 2 t^2)
\nonumber \\
&\quad + s^2 t (s + t)
\Big].
\end{align}

The thermally-averaged annihilation cross-section reads
\beq
\langle \sigma v_{\text{rel}} \rangle_{\chi\Bar{\chi}\to W^+ W^-}
&=
\frac{2\alpha g^2 m_\chi^6}{3\Lambda^4 m_W^4}\langle v_{\text{rel}}^2 \rangle \, .
\eeq
\subsubsection{Electric Dipole Moment (EDM)}
\label{app:annihilation CS EDM}
If kinematically allowed, 
$\chi$ can annihilate into all the SM leptons and quarks, as well as into the 
$W$ boson and photons. The differential cross-section for DM annihilation into SM fermions ($f$)
in the CM frame is given by
\begin{equation}
\frac{d\sigma}{d\Omega}\bigg|_{\chi\Bar{\chi}\to f\bar{f}}
= \frac{1}{64\pi^2 s}
\sqrt{\frac{s - 4 m_f^2}{s - 4 m_\chi^2}}
|\overline{\mathcal{M}}_{f \bar{f}}|^2 
\end{equation}
where the spin-averaged squared amplitude reads
\begin{equation}
\begin{aligned}
|\overline{\mathcal{M}}_{f \bar{f}}|^2
&= \frac{16C_F e^2 g^2}{\Lambda^2 s}
\Big[
2 m_\chi^2 t + m_f^2 (s + 2t - 2 m_\chi^2)
- m_\chi^4 \\
&\quad
- m_f^4
- t (s + t)
\Big] ,
\end{aligned}
\end{equation}
in terms of the Mandelstam variables $s$, $t$, $u$ and the SM fermion masses $m_f$.

In the NR limit, the thermally averaged annihilation cross-section can be expanded in the relative velocity 
$v_{\text{rel}}$. At leading order, one obtains
% \begin{equation}
% \begin{aligned}
% \langle \sigma v_{\text{rel}} \rangle_{\chi \chi \to f \bar{f}}&=
% \frac{\alpha g^2 }{2\Lambda^2}\sqrt{1 - \frac{m_f^2}{m_\chi^2}}\\
% &\quad\times\Bigg[1-\frac{1}{3}\left(1 - \frac{m_f^2}{m_\chi^2}\right)\Bigg]
% \langle v_{\text{rel}}^2 \rangle.
% \end{aligned}
% \end{equation}

\begin{equation}
\begin{aligned}
\langle \sigma v_{\text{rel}} \rangle_{\chi \Bar{\chi} \to f \bar{f}}&=
\frac{\alpha C_F g^2 }{3\Lambda^2}\sqrt{1 - \frac{m_f^2}{m_\chi^2}}\left(1 + \frac{m_f^2}{2 m_\chi^2}\right)
\langle v_{\text{rel}}^2 \rangle.
\end{aligned}
\end{equation}
Similarly, the differential cross-section for DM annihilation into $W^+W^-$ in the CM frame is given by
\begin{equation}
\frac{d\sigma}{d\Omega}\bigg|_{\chi\Bar{\chi}\to W^+ W^-}=\frac{1}{64\pi^2 s}
\sqrt{\frac{s- 4 m_W^2 }{s - 4 m_\chi^2}}
\,|\overline{\mathcal{M}}_{W^+ W^-}|^2 ,
\end{equation}
where,
\begin{equation}
\begin{split}
|\overline{\mathcal{M}}_{W^+W^-}|^2
=&\,
\frac{e^{2} g^{2}}
     {2 \Lambda^{2} m_{W}^{4} s}
\Big[
4 m_{\chi}^{4}
\left(
12 m_{W}^{4}
-4 m_{W}^{2} s
+s^{2}
\right)
\\
&+
4 m_{\chi}^{2}
\Big(
24 m_{W}^{6}
+12 m_{W}^{4}(s-2t)
\\
&\qquad
-2 m_{W}^{2}s(s-4t)
-s^{2}(s+2t)
\Big)
\\
&+
48 m_{W}^{8}
-32 m_{W}^{6}(2s+3t)
\\
&+
16 m_{W}^{4}t(5s+3t)
-8 m_{W}^{2}st(3s+2t)
\\
&+
s^{2}(s+2t)^{2}
\Big].
\end{split}
\end{equation}
In the NR limit, the corresponding thermally averaged annihilation cross-section reads
\beq
\langle \sigma v_{\text{rel}} \rangle_{\chi\Bar{\chi}\to W^+ W^-}
&=
\frac{\alpha g^2 m_\chi^4}{3\Lambda^2 m_W^4}\langle v_{\text{rel}}^2 \rangle\,.
\eeq
DM can also annihilate into photons. The differential cross-section is
\begin{equation}
\frac{d\sigma}{d\Omega}\bigg|_{\chi\Bar{\chi}\to \gamma\gamma}
= \frac{1}{64\pi^2 s}
\sqrt{\frac{s}{s - 4 m_\chi^2}}
|\overline{\mathcal{M}}{\gamma\gamma}|^2 
\end{equation}
with
\begin{equation}
\begin{aligned}
|\overline{\mathcal{M}}_{\gamma\gamma}|^2
=&\,
\frac{64 g^4}
{\Lambda^{4}(m_\chi^{2}-t)(m_\chi^{2}-s-t)}
\\
&\times
\Big[
4m_\chi^{6}(s+t)
-t^{2}(s+t)^{2}
\\
&\qquad
+4m_\chi^{2}t(s+t)^{2}
-m_\chi^{4}(s^{2}+10st+6t^{2})-m_\chi^{8}\Big].
\end{aligned}
\end{equation}
In the non-relativistic limit,
\begin{equation}
\langle \sigma v_{\rm rel} \rangle_{\chi\Bar{\chi}\to \gamma\gamma}=
\frac{g^4 m_\chi^2}{\pi \Lambda^4}
\left(8 + 5 \langle v_{\rm rel}^2 \rangle \right) .
\end{equation}
 \subsubsection{Magnetic Dipole Moment (MDM)}
 \label{app:annihilation CS MDM}
 Whenever kinematically accessible, $\chi$ can annihilate into all the SM leptons and quarks, as well as into gauge boson final states such as $W^+W^-$ and $\gamma\gamma$. The differential cross-section for annihilation into SM fermions (f) in the CM frame is given by
\begin{equation}
\frac{d\sigma}{d\Omega}\bigg|_{\chi\Bar{\chi}\to f\bar{f}}
= \frac{1}{64\pi^2 s}
\sqrt{\frac{s - 4 m_f^2}{s - 4 m\chi^2}}
|\overline{\mathcal{M}}_{f \bar{f}}|^2 
\end{equation}
where the corresponding spin-averaged squared amplitude takes the form
\begin{equation}
\begin{aligned}
|\overline{\mathcal{M}}_{f \bar{f}}|^2
&= \frac{16 C_F e^2 g^2}{\Lambda^2 s}
\Big[
2 m_\chi^2 t + 2 m_\chi^2 s + 
m_f^2 (s + 2 t + 2 m_\chi^2)  \\
&\quad
- m_\chi^4 - m_f^4 - t (s + t)
\Big] \,.
\end{aligned}
\end{equation}
Expanding in the relative velocity $v_{\rm rel}$, the thermally averaged annihilation cross-section in the NR regime can be written as a velocity expansion. In the limit 
$m_f\rightarrow 0$, the leading contribution is
\begin{equation}
\langle \sigma v_{\rm rel} \rangle_{\chi \Bar{\chi} \to f \bar{f}}=
\frac{4\alpha C_F g^2}{\Lambda^2}
\left(1 - \frac{1}{24}\langle v_{\rm rel}^2 \rangle \right).
\end{equation}
For annihilation into $W^+W^-$, the differential cross-section in the CM frame is given by
\begin{equation}
\frac{d\sigma}{d\Omega}\bigg|_{\chi\Bar{\chi}\to W^+W^-}
= \frac{1}{64\pi^2 s}
\sqrt{\frac{s - 4 m_W^2}{s - 4 m\chi^2}}
|\overline{\mathcal{M}}_{W^+W^-}|^2 
\end{equation}
with the squared amplitude
\begin{equation}
\begin{aligned}
|\overline{\mathcal{M}}_{W^+W^-}|^2
&= \frac{e^{2} g^{2}}{2 \Lambda^{2} m_W^{4} s}
\Big[
4 m_\chi^{4} (12 m_W^{4} - 4 m_W^{2} s + s^{2}) \\
&\quad
- 8 m_\chi^{2} \big(12 m_W^{6} + 4 m_W^{4} (7 s + 3 t) \\
&\quad
- m_W^{2} s (7 s + 4 t) + s^{2} t \big) + 48 m_W^{8} \\
&\quad
- 32 m_W^{6} (2 s + 3 t) + 16 m_W^{4} t (5 s + 3 t) \\
&\quad
- 8 m_W^{2} s t (3 s + 2 t) 
+ s^{2} (s + 2 t)^{2}
\Big] .
\end{aligned}
\end{equation}
The associated thermally averaged cross-section in the NR limit is given by
\begin{equation}
\langle \sigma v_{\rm rel} \rangle_{\chi\Bar{\chi}\to W^+ W^-}
= \frac{4 \alpha g^2 m_\chi^4}{\Lambda^2 m_W^4}
\left(1 + \frac{\langle v_{\rm rel}^2\rangle}{3}\right) .
\end{equation}
In addition, annihilation into photons proceeds via 
$t$- and $u$-channel contributions. The corresponding differential cross-section is
\begin{equation}
\frac{d\sigma}{d\Omega}\bigg|_{\chi\Bar{\chi}\to \gamma \gamma}
= \frac{1}{64\pi^2 s}
\sqrt{\frac{s}{s - 4 m\chi^2}}
|\overline{\mathcal{M}}_{\gamma\gamma}|^2 
\end{equation}
where the squared matrix element is
\begin{equation}
\begin{aligned}
|\overline{\mathcal{M}}_{\gamma\gamma}|^2
=&\,
\frac{64 g^4}
{\Lambda^{4}(m_\chi^{2}-t)(m_\chi^{2}-s-t)}
\\
&\times
\Big[
4m_\chi^{6}(s+t)
-t^{2}(s+t)^{2}
\\
&\quad
+4m_\chi^{2}t(s+t)^{2}
-m_\chi^{4}(s^{2}+10st+6t^{2})
-m_\chi^{8}
\Big].
\end{aligned}
\end{equation}
In the NR limit, the thermally averaged annihilation cross-section takes the form
\begin{equation}
\langle \sigma v_{\rm rel} \rangle_{\chi\Bar{\chi}\to \gamma \gamma}
=\frac{g^4 m_\chi^2}{\pi \Lambda^4}
\left(8 + 5 \langle v_{\rm rel}^2 \rangle\right) .
\end{equation}

\subsubsection{Charge Radius (CR)}
$\chi$ can annihilate into all the SM leptons and quarks. The spin-averaged squared
amplitude is given by,
\beq
|\overline{\mathcal{M}}_{f \bar{f}}|^2&= \frac{2 e^2 g^2 C_F}{\Lambda^4}
\Big[2 m_\chi^4+ 4 m_\chi^2 (m_f^2 - t) \\
& \hspace{1.0cm} + 2 m_f^4 - 4 m_f^2 t + s^2+ 2 s t+ 2 t^2\Big]
\eeq
In the NR regime, the thermally averaged annihilation cross-section can be expanded in powers of the relative velocity $v_{\rm rel}$. In the limit $m_f \rightarrow 0$, the leading contribution takes the form
\begin{equation}
\langle \sigma v_{\rm rel} \rangle_{\chi \Bar{\chi} \to f \bar{f}}=
\frac{\alpha g^2 C_F m_\chi^2}{\Lambda^4}
\left(4 + \frac{7}{6}\langle v_{\rm rel}^2 \rangle \right) .
\end{equation}
For annihilation into 
$W^+W^-$, the differential cross-section in the CM frame is given by
\begin{equation}
\frac{d\sigma}{d\Omega}\bigg|_{\chi\Bar{\chi}\to W^+W^-}
= \frac{1}{64\pi^2 s}
\sqrt{\frac{s - 4 m_W^2}{s - 4 m\chi^2}}
|\overline{\mathcal{M}}_{W^+W^-}|^2 ,
\end{equation}
with the squared amplitude
\beq
\overline{|\mathcal{M}}_{W^+W^-}|^2=
-\frac{e^2 g^2}{2\Lambda^4 m_W^4}
\Big[
m_\chi^4(12m_W^4-4m_W^2s+s^2)
\\
-2m_\chi^2\Big(
12m_W^6
+4m_W^4(7s+3t)
-m_W^2s(7s+4t)
+s^2t
\Big)
\\
+12m_W^8
-4m_W^6(s+6t)
+m_W^4(17s^2+20st+12t^2)
\\
-2m_W^2s(2s^2+3st+2t^2)
+s^2t(s+t)
\Big].
\eeq
The thermally averaged cross-section in the NR limit is given by
\begin{equation}
\langle \sigma v_{\rm rel} \rangle_{\chi\Bar{\chi}       \to W^+W^-}
=
\frac{\,\alpha\, g^2\, m_\chi^6}
{\,\Lambda^4\, m_W^4}
\left(4 + \frac{19}{6}\langle v_{\rm rel}^2 \rangle \right) .
\end{equation}
\bibliography{refs_DM}
\end{document}